\documentclass[11pt]{article}

\usepackage{amsmath,amssymb,bbm}
\usepackage{cancel}
 
\def\dblone{\hbox{$1\hskip -1.2pt\vrule depth 0pt height 1.6ex width 0.7pt
                  \vrule depth 0pt height 0.3pt width 0.12em$}}

\begin{document}
\pagestyle{empty}
\rightline{LTH 1164}
\vskip 1.5 true cm  
\begin{center}  
{\large Five-dimensional vector multiplets in arbitrary signature}\\[.5em]
\vskip 1.0 true cm   
{L.~Gall$^{1}$ and T.~Mohaupt$^{1}$} \\[5pt] 
$^1${Department of Mathematical Sciences\\ 
University of Liverpool\\
Peach Street \\
Liverpool L69 7ZL, UK\\[2ex]  
Louis.Gall@liverpool.ac.uk,
Thomas.Mohaupt@liv.ac.uk \\[1em]
}
June 19, 2018, 
revised: August 24, 2018
\end{center}
\vskip 1.0 true cm  
\baselineskip=18pt  
\begin{abstract}  
\noindent  
We start developing a formalism which allows to construct
supersymmetric theories systematically across space-time 
signatures. Our construction uses a complex form of the 
supersymmetry algebra, which is obtained by doubling the spinor
representation. This allows one to partially disentangle the Lorentz and
R-symmetry group  and generalizes symplectic Majorana spinors.
For the case where the spinor representation is 
complex-irreducible, the R-symmetry only acts on an internal 
multiplicity space, and we show that the connected groups which
occur are $SO(2), SO_0(1,1), SU(2)$ and $SU(1,1)$.

As an application we construct
the off-shell supersymmetry transformations and supersymmetric 
Lagrangians for five-dimensional vector multiplets in arbitrary signature
$(t,s)$. In this case the R-symmetry groups are $SU(2)$ or $SU(1,1)$,
depending on whether the spinor representation carries a quaternionic
or para-quaternionic structure. In Euclidean signature the scalar and
vector kinetic terms differ by a relative sign, which is consistent
with previous results in the literature and shows that this sign flip
is an inevitable consequence of the Euclidean supersymmetry algebra.

\end{abstract}


\newpage
 \pagestyle{plain}
\tableofcontents

\section{Introduction and Summary of Results}

Non-Lorentzian space-time signatures are relevant for a variety of
reasons. For Euclidean signature this is obvious, since the study
of non-perturbative effects, such as instantons, makes use of 
the Euclidean functional integral formalism. Moreover, the stationary
sectors of Lorentzian theories can be studied through effective
Euclidean theories obtained by dimensional reduction over time. 
In quantum gravity one might also ask whether the Lorentzian signature
of space-time is to be accepted as `given,' or whether it needs to
be explained, for example within a framework where signature change
occurs dynamically. In string theory the study of the web of dualities
connecting different types of string theories naturally leads to the
inclusion of exotic, multi-time signatures once T-duality
transformations with respect to time are 
considered \cite{Hull:1998vg,Hull:1998ym,Dijkgraaf:2016lym}. 
To facilitate such studies, a systematic way to 
construct and relate theories, in particular supersymmetric theories,
across space-time signatures is needed. One common
approach to changing space-time signature, is the ad hoc flipping of signs
and insertion of factors of $i$. A more systematic and well
developed way to obtain Euclidean theories is to start with a
supersymmetric theory in Lorentzian signature and then to
dimensionally reduce over time \cite{Hull:1998br,Cremmer:1998em,Cortes:2003zd,Cortes:2005uq,Cortes:2009cs,Cortes:2015wca,Gutowski:2012yb,Sabra:2015tsa,Sabra:2016abd,Sabra:2017xvx,deWit:2017cle}. While this guarantees one finds 
a theory invariant under Euclidean supersymmetry, it excludes theories
that cannot be obtained by dimensional reduction. Moreover, only in a
few cases \cite{Cortes:2003zd,deWit:2017cle}
have these reductions been carried out in an off-shell
formulation and have fully included the fermions. In most cases attention
has been restricted to on-shell formulations and to bosonic terms,
while the fermions have been neglected or only been considered
through  the Killing spinor
equations evaluated on bosonic backgrounds. In \cite{Sabra:2017xvx}
analytic continuation of the Killing spinor equations were used to
obtain the supersymmetry variations of the fermions, and, by 
imposing closure of the algebra, the bosonic terms of the on-shell
Lagrangian for five-dimensional vector multiplets coupled to 
supergravity for all signatures. 

A systematic construction should start with the supersymmetry algebra,
then where possible construct an off-shell representation of the
algebra on fields, and only then proceed to invariant Lagrangians. The
natural way to obtain a universal construction that allows to treat
all signatures in a given dimension simultanously is to work with 
the complex form of the supersymmetry algebra, to construct its
representations on complexified fields, and finally a `holomorpic
master Lagrangian.' To specialize to particular signatures one
imposes suitable reality conditions and obtains the corresponding 
field representations and Lagrangians. In this paper we start
developing such a formalism and demonstrate its viability by 
applying it to the case of rigid  five-dimensional off-shell vector  
multiplets. Our approach makes use of the relation between Poincar\'e
Lie superalgebras and real, symmetric, Spin$_0$-equivariant
vector-valued bilinear forms. This allows us to use results from the
mathematical literature, in particular the classification Poincar\'e
Lie superalgebras in arbitrary dimension and signature \cite{Alekseevsky:1997}. While in
spirit our approach is similar to the treatment of maximal
supergravity theories in dimensions 10 and 11 in
\cite{Bergshoeff:2000qu,Bergshoeff:2007cg}, one  difference is that we do not need to assume that the
supersymmetry algebra arises as a contraction of an ortho-symplectic
Lie superalgebra. This is important since it is not clear a priori whether all
Poincar\'e Lie superalgebras (in all signatures) can be obtained as 
contractions. BPS-charges, also called polyvector charges, or, by
abuse of terminology, central charges, which in the ortho-symplectic
framework arise through different contractions, can be added to our
formalism using the work of \cite{Alekseevsky:2003vw}, though we will
leave this 
aspect
to future work.

The core piece of formalism that we develop in the first part of this
paper is what we call the doubled spinor module. This accounts for
the doubling, or complexification, of the supercharges and fermionic
fields that characterizes the complex form of the theory. As we will
explain in more detail in the paper, the essential part of defining a
supersymmetry algebra is to choose a so-called  admissible bilinear
form on the spinor module ${\bf S}$ of the Poincar\'e Lie superalgebra
$\mathfrak{sp}(V) = V + \mathfrak{so}(V) + {\bf S}$, where
$V=\mathbb{R}^{t,s}$ is a space-time with signature
$(t,s)$. In short,
the admissible bilinear form determines the superbracket between the
supercharges of the theory. The complex form of the theory is defined
using the doubled spinor space ${\bf S}\oplus {\bf S}$, which is 
equipped with a {\em complex}, symmetric vector-valued bilinear form, 
which is equivariant with respect to the complex Spin group, and thus
gives rise to a superbracket for the complex form of the supersymmetry
algebra.  Real forms are obtained
by picking a Spin$(V)$-invariant real structure on ${\bf S}
\oplus {\bf S}$ and restricting to the real points $({\bf S} \oplus
{\bf S})^\rho \simeq {\bf S}$. The restriction of the complex bilinear
form gives rise to the real bilinear form associated to the real
superbracket. A well known example of reformulating spinors
by first doubling and then imposing a reality condition are symplectic
Majorana
spinors, which are naturally included in our formalism.
One advantage of the doubling is that part
of the R-symmetry group becomes manifest when writing the 
doubled spinor module as a complex tensor product, ${\bf S} \oplus
{\bf S} \simeq {\bf S} \otimes \mathbb{C}^2$. In this product form, the
Lorentz Lie algebra only acts on the first factor, whereas the
R-symmetry acts in general on both ${\bf S}$ and on the multiplicity 
space $\mathbb{C}^2$. In the simplest case, where ${\bf S}$ is complex
irreducible, that is for Dirac spinors on odd dimensions, 
Schur's lemma implies that the R-symmetry group
can be taken to only act on $\mathbb{C}^2$. In this case the R-symmetry 
group of the complex theory is either $O(2,\mathbb{C})$ or
$Sp(2,\mathbb{C})$,  depending on whether the complex bilinear form is symmetric or
antisymmetric when restricted to $\mathbb{C}^2$. Since the reality
condition selects a real form of the complex R-symmetry group, it
follows that the connected part of the real R-symmetry group must be
$SO(2), SO_0(1,1), USp(2) \simeq SU(2)$ or $Sp(2,\mathbb{R}) \simeq
SU(1,1)$. In the general case, where ${\bf S}$ is complex reducible, 
these groups are part of a larger R-symmetry group. In this paper we 
will work out the R-symmetry groups for supersymmetry algebras based on
complex irreducible spinor modules, while the general case will be 
presented in a separate paper \cite{R-symmetry}.

Having developed the general formalism to this point, we then
specialize to five dimensions and to vector multiplets. 
In five dimensions the complex spinor module $\mathbb{S}$, that is
the representation by Dirac spinors, is both complex and real
irreducible in Lorentzian signature and defines the unique minimal
supersymmetry algebra. Counting supercharges in multiples of the
minimal four-dimensional Lorentzian signature supersymmetry algebra,
we refer to this algebra as the five-dimensional ${\cal N}=2$
supersymmetry algebra. While for signatures $(2,3)$ and $(3,2)$ Dirac
spinors are real reducible, that is, one can define Majorana spinors,
we show that there is not corresponding ${\cal N}=1$ supersymmetry
algebra with four real supercharges, since the superbracket is
trivial. Thus there is a unique minimal supersymmetry algebra with
eight real supercharges for all five-dimensional signatures. 
We then carry out the programme outlined above and obtain
the off-shell field representations and invariant Lagrangians of 
vector multiplets for all signatures $(5,0), \ldots, (0,5)$. From our
results it is manifest that theories which are space-time mirrors, 
that is related by 
$t \leftrightarrow s$, are physically equivalent in the sense that all sign
flips and factors $i$ are determined by going between a mostly plus and a
mostly minus convention for the space-time metric, plus taking into
account the signature dependence of the reality properties of spinor
bilinears. Throughout the paper we adopt the convention of referring 
to the smaller of the numbers $t,s$ as time, so that there are at most
two time-like dimensions. In the Euclidean signatures $(0,5)$ and
$(5,0)$ we find that the scalar and vector terms always come with a
relative sign. This confirms the indirect arguments given in \cite{Sabra:2016abd}
and shows that this relative sign follows inevitably from the
Euclidean supersymmetry algebra. Note that this is not clear a priori.
In particular, it has been argued that a similar sign flip for four-dimensional vector
multiplets can be removed by a field redefinition 
\cite{Sabra:2015tsa,Sabra:2016abd}. We also compare our results
to those of \cite{Sabra:2017xvx}, where the bosonic on-shell
Lagrangians for five-dimensional vector multiplets coupled to supergravity
have been obtained for all signature by analytic continuation of 
the Killing spinor equations of the Lorentzian theory. We find
that all signature dependent relative signs agree.


The outline of this paper is as follows. In Section 2 we first present the 
background on Clifford algebras and spinors needed to make this
paper self-contained. Then we review the relation between
supersymmetry algebras and vector-valued bilinear forms. 
We introduce the doubled spinor construction and analyse 
which R-symmetry groups can occur for supersymmetry algebras based on
complex irreducible spinors. 
In Section 3 we turn to minimal supersymmetry in five dimensions,
where the relevant spinor representation ${\bf S}$ is the one by Dirac
spinors denoted $\mathbb{S}$. 
We show explicitly how the superbrackets formulated on the complex
spinor module $\mathbb{S}$ 
and on the doubled spinor module 
$\mathbb{S} \otimes \mathbb{C}^2$ are related, and show that while
Majorana spinors can be defined for some signatures, the corresponding
${\cal N}=1$ supersymmetry algebras are trivial. 
In Section 4 we construct an off-shell representation 
of the complex supersymmetry algebra by vector multiplets,
and a corresponding invariant off-shell Lagrangian. By imposing
reality conditions, we obtain the transformations and Lagrangians
for all signatures from 
$(0,5)$ to $(5,0)$.  In Section 5 we give a a brief outlook onto open 
problems and future 
directions. Some material has been relegated to appendices in order
not to avoid breaking the flow of the presentation. While we suppress spinor 
indices most of the time, we need to give the details which allow us
to translate between vector-valued bilinear forms and supercharge
anticommutators. This is done in Appendix A. Appendices B and C
contain the relevant background on para-quaternions and quaternions.

\section{Supersymmetry and bilinear forms}

In this section we work in arbitrary dimension and signature. After
providing the background on Clifford algebras, supersymmetry algebras,
and admissible bilinear forms needed to make this article
self-contained, we introduce the formalism based of doubled 
spinors, which  allows to define a complex form of the supersymmetry 
algebra and then to specialize to any signature by
imposing reality conditions. We review the useful concepts of the
Schur algebra and Schur group, and investigate how the choice of
a reality condition determines the R-symmetry group of the real
superalgebra.

\subsection{Clifford algebras and spinors \label{Sect:Cliff+Spin}}

We work on a `space-time' $V=\mathbb{R}^{t,s}$ of signature $(t,s)$ with metric
\[
\eta = \mbox{diag}( \underbrace{-1 \;, -1 \ldots \;,}_{t}
\underbrace{+1\;, +1 \;, \ldots }_{s} ) 
\]
and isometry Lie algebra  $\mathfrak{p}(V) = V + \mathfrak{so}(V)$,
where $V=\mathbb{R}^{t,s}$ are the translations\footnote{We do not
  distinguish by notation between the vector space with inner product
$V=\mathbb{R}^{t,s}=(\mathbb{R}^{t+s}, \eta)$, the associated affine space, and
the  translation group $(\mathbb{R}^{t+s}, +)$ acting on the affine space.} and 
$\mathfrak{so}(V) = \mathfrak{so}(t,s)$ the linear transformations
preserving $\eta$. 
Space-time indices are denoted $\mu, \nu=1,\ldots, t+s$. 
The real Clifford algebra $Cl(V)=Cl_{t,s}$ has generators $\gamma_\mu$,
with defining relations 
\begin{equation}
	\{ \gamma_{\mu}, \gamma_{\nu} \} = 2 \eta_{\mu \nu} \;.
\end{equation} 
Lorentz indices on Clifford generators are raised and lowered using $\eta$, that is
$\gamma^\mu = \eta^{\mu \nu} \gamma_\nu$. 
Our conventions are the
same as in \cite{Cortes:2003zd}. Note that they differ from the conventions in
\cite{SpinGeometry}
by a relative sign in the definining relation
and another sign in the definition of $\eta$. The combined effect of
both relative signs is that 
$Cl_{t,s}$ refers to the same associative real algebra.
While our notation is
adapted to the `mostly plus' convention for Minkowski space-time,
we will interpret $\mbox{min}\{t,s \}$ as the number of time-like
directions, because we expect that theories related by an overall 
sign change, $t\leftrightarrow s$ of the metric are physically equivalent.
Thus, for example, in four and five dimensions there are
at most two time-like directions.

We use a representation of the Clifford generators in terms of
complex square matrices of size $2^{[(t+s)/2]}$, with the following
Hermiticity properties:
\[
(\gamma^\mu)^\dagger = \left\{ \begin{array}{ll}
-\gamma^\mu \;, & \mbox{for}\;\mu=1, \ldots, t \;, \\
\gamma^\mu \;, & \mbox{for}\;\mu=t+1, \ldots, t+s \;.\\
\end{array}\right.
\]
This fixes the $\gamma$-matrices up to unitary equivalence. The
$\gamma$-matrices provide an irreducible representation
of the  complex
Clifford algebra $\mathbb{C}l_{t+s} = Cl_{t,s}\otimes \mathbb{C}$.
By restriction to a representation of the real group $\mbox{Spin}(t,s)\subset
Cl_{t,s} \subset \mathbb{C}l_{t+s}$ one obtains a complex spinor
representation of the Lorentz group, the representation 
by Dirac spinors. The corresponding representation space will be
referred to as the complex spinor module $\mathbb{S}$. 
Dirac spinors provide an irreducible complex 
$\mbox{Spin}(t,s)$- representation for odd $t+s$, while for even $t+s$ Dirac spinors
decompose into Weyl spinors, $\mathbb{S}=\mathbb{S}_+\oplus
\mathbb{S}_-$.

The $\gamma$-matrices can be related to the Hermitian conjugate,
complex conjugate and transposed matrices by relations of the form 
\cite{Cortes:2003zd}\footnote{See also \cite{VanProeyen:1999ni} for 
a comprehensive review. Note that the parameters $\epsilon$ and $\eta$ used
there are related to our parameters by $\sigma = - \epsilon$ and $\tau
= - \eta$.} 
\begin{eqnarray}
(\gamma^\mu)^\dagger &=& (-1)^t A \gamma^\mu A^{-1}\;, \\
(\gamma^\mu)^* &=& (-1)^t \tau  B \gamma^\mu B^{-1}\;, \\
(\gamma^\mu)^T &=&  \tau  C \gamma^\mu C^{-1}\;, 
\end{eqnarray}
where $\tau=\pm 1$.
For $A$ we make the explicit choice
\[
A = \gamma_1 \cdots \gamma_t \;,
\]
which implies
\[
A^{\dagger} = (-1)^{\frac{t(t+1)}{2}} A \;,\;\;\;
A^{-1} = (-1)^t \gamma_t \cdots \gamma_1 \;.
\]
The matrix $A$ generalizes the `$\Gamma_0$-matrix' familar from 
Lorentz signature and enters into the definition of Dirac spinor bilinears,
and of the Dirac conjugate of a spinor, $\bar{\psi}^{(D)} =
\psi^\dagger A$. The matrix $C$ is the charge conjugation matrix, which relates
particle and anti-particle states, and defines the Majorana conjugate
$\bar{\psi} = \psi^T C$ 
of  spinors. It can be chosen to satisfy (see for example \cite{VanProeyen:1999ni})
\[
C^{-1} = C = C^\dagger \;.
\]
$C$ is either symmetric or
antisymmetric,
\[
C^T = \sigma C \;,
\]
where $\sigma=\pm 1$. Which values of $\sigma$ and $\tau$
can occur depends on the dimension. For each value of $\tau$ there
is always precisely one corresponding value of $\sigma$.
In even dimensions, $t+s=2n$, both $\tau=1$ and $\tau=-1$ are
possible, 
and  there are two inequivalent 
choices for $C$, denoted $C_{-\tau}$.
In odd dimensions, $t+s=2n+1$,  the matrix $C$ is unique up to
equivalence, and  $\tau=(-1)^{(t+s-1)/2)}$.

Given $A$ and $C$, the matrix $B$ can be chosen as
\[
B = (CA^{-1})^T \;.
\]
Then $B$ is unitary, $B^\dagger B = \dblone$, and
\[
BB^* = \pm \dblone \;,
\]
where the sign depends on the signature, and, in even
dimensions, on the choice of the charge conjugation matrix $C=C_\pm$.
In signatures where it is possible to choose 
$BB^*=+\dblone$, the matrix $B$ can be used to
impose a reality condition and to define Majorana
spinors. Mathematically, $B$ defines a family of real structures on the complex
spinor module $\mathbb{S}$, 
\[
\rho^{(\alpha)}_{\mathbb{S}} (\lambda) = \alpha^* B^* \lambda^*
\;,\;\;
(\rho^{(\alpha)}_{\mathbb{S}})^2 = \dblone \;,
\]
where $\alpha \in \mathbb{C}$ is a phase, $|\alpha|=1$. The presence of
this phase reflects that multiplying $B$ by a phase does not change
the relations $BB^\dagger = \dblone$ and $BB^*=\pm \dblone$. We
have made an explicit choice for $B$ in terms of $A$ and $C$, but 
will use the freedom represented by $\alpha$ later to choose 
our reality conditions such that the resulting formulae are uniform
for  all signatures. The real structure 
defined by $B$ is Spin$_0$-invariant:\footnote{That is, it is
  invariant under the Lie algebra $\mathfrak{so}(V)$, and hence under
the connected component of the Spin group.}
\[
\rho^{(\alpha)}_{\mathbb{S}} (\gamma^\mu \lambda) =
\alpha^* B^* (\gamma^\mu)^* \lambda^* =
(-1)^t \tau \gamma^\mu B^* \lambda^*  \;,
\]
and therefore
\[
\rho^{(\alpha)}_{\mathbb{S}} (\gamma^{\mu\nu} \lambda) =
\alpha^* B^* (\gamma^{\mu\nu})^* \lambda^* =
(-1)^{2t} \tau^2 \gamma^{\mu\nu} \alpha^*  B^* \lambda^* =
\gamma^{\mu\nu} \rho^{(\alpha)}_{\mathbb{S}}(\lambda)  \;.
\]
Majorana spinors $\lambda \in \mathbb{S}_M\subset \mathbb{S}$ are
by definition the spinors which are invariant under the real
structure, 
\[
\rho^{(\alpha)}_{\mathbb{S}} (\lambda) = \lambda \Leftrightarrow
\lambda^* = \alpha B \lambda \;.
\]

In  signatures where $BB^*=-\dblone$, the matrix $B$ defines a
family of 
Spin$_0$-invariant quaternionic structures, 
\[
j^{(\alpha)}_{\mathbb{S}} (\lambda) = \alpha^* B^* \lambda^* \;,\;\;\;
(j^{(\alpha)}_{\mathbb{S}})^2 = - \dblone \;.
\]
The Spin$_0$-invariance holds for the same reason as for
the real structure: we have
$j^{(\alpha)}_{\mathbb{S}}(\gamma^\mu \lambda) = \pm \gamma^\mu
j^{(\alpha)}_{\mathbb{S}}(\lambda)$, and therefore
$j^{(\alpha)}_{\mathbb{S}}(\gamma^{\mu\nu} \lambda) = \gamma^{\mu\nu}
j^{(\alpha)}_{\mathbb{S}}(\lambda)$. While a quaternionic structure
cannot be used to define a reality condition on $\mathbb{S}$ itself,
it can be used to impose a reality condition on complex spaces 
of the form $\mathbb{S}\otimes V_{\mathbb{C}}$, where $V_{\mathbb{C}}$
is a complex vector space equipped with a quaternionic structure. 
Symplectic Majorana spinors are one example of such a construction and
will be discussed later.

We remark that the complex spinor module $\mathbb{S}\simeq
\mathbb{C}^{2n}$  can always either be viewed as a vector space
$\mathbb{S}\simeq \mathbb{H}^n$ over the skew field $\mathbb{H}$ 
of quaternions, or as a module $\mathbb{H}'^n$ over the ring of
para-quaternions $\mathbb{H'}$.\footnote{We refer to the appendices for brief reviews of the relevant facts
about quaternions and para-quaternions.  
}
The reason is that the Spin$_0$-invariant endomorphism defined by
\[
J = \left\{  \begin{array}{ll}
j^{(\alpha)}_\mathbb{S}\;, & \mbox{if}\;\;\mathbb{S}\;\;\mbox{has an invariant
               quaterionic structure}\;, \\
\rho^{(\alpha)}_\mathbb{S}\;, & \mbox{if}\;\;\mathbb{S}\;\;\mbox{has an invariant
               real structure} \;,\\
\end{array} \right.
\]
anti-commutes with the natural complex 
structure $I$, which acts through multiplication by the complex unit,
$I:\lambda \mapsto i \lambda$. Therefore $K=IJ=-JI$ is another 
Spin$_0$-invariant endomorphism which anti-commutes with $I$ and $J$.
If $BB^*=\varepsilon \dblone$, $\varepsilon =\pm 1$
then the three endomorphisms $I,J,K$ satisfy
\[
I^2 = \varepsilon J^2 = \varepsilon K^2 = - \dblone \;,\;\;\;
I,J,K \;\;\mbox{pairwise anticommuting,}
\]
which is the quaternionic algebra for $\varepsilon=1$ and the
para-quaternionic
algebra for $\varepsilon=-1$.

\subsection{Bilinear Forms, Supersymmetry, and R-Symmetry Groups}

\subsubsection{Supersymmetry algebras and bilinear forms}

Supersymmetry extends the Poincar\'e Lie algebra $\mathfrak{p}(V) = V
+ \mathfrak{so}(V)$ to a Poincar\'e Lie superalgebra $\mathfrak{sp}(V)
= \mathfrak{g}_0 + \mathfrak{g}_1$, where the even part
$\mathfrak{g}_0$ is the Poincar\'e Lie algebra, and the odd part 
$\mathfrak{g}_1$ is a spinor module 
${\bf S}$ of $\mathfrak{so}(V)$,
which may be reducible or irreducible. The Poincar\'e Lie superalgebra
is completely determined once we specify the supersymmetry algebra 
(supertranslation algebra) 
\begin{equation}
\label{susy_algebra_generic}
\{ Q_\alpha, Q_\beta \} = K^\mu_{\;\;\alpha \beta} P_\mu \;,
\end{equation}
where the translation generators $P_\mu$ span $V$, while the 
supercharges $Q_\alpha$ span ${\bf S}$. 
Since the bracket $\{ \cdot, \cdot \}$ is symmetric, the structure constants
$K^\mu_{\;\;\alpha \beta}$ can be interpreted as representing a real
Spin$_0$-equivariant symmetric vector-valued bilinear form on the spinor 
module ${\bf S}$:
\[
K \;: \mbox{Sym} ({\bf S} \otimes {\bf S} ) \rightarrow
\mathbb{R}^{t,s} \;:\;\;\; (\lambda, \chi) \mapsto K^\mu(\lambda,\chi)
= K^\mu(\chi, \lambda)\;.
\]
Conversely, any real Spin$_0$-equivariant, symmetric vector-valued bilinear
form defines a superbracket. The anti-commutation relations
(\ref{susy_algebra_generic}) are obtained
by expanding spinors in the basis provided by the supercharges $Q_\alpha$: 
\[
\{ \lambda^\alpha Q_\alpha, \chi^\beta Q_\beta \} =
K^\mu (\lambda, \chi) P_\mu = \lambda^\alpha \chi^\beta
K^\mu_{\;\;\alpha \beta} P_\mu \;.
\]
Note that in this approach there is no need to explicitly check 
the super Jacobi identity, because it holds automatically.
It was shown in \cite{Alekseevsky:1997} how all supersymmetry
algebras in all dimensions and signatures can be constructed from 
vector-valued bilinear forms. In the following we draw on results of
this work to explain how such vector-valued bilinear forms can be
constructed using so-called admissible bilinear forms. We will not
consider central charges and other BPS charges, which correspond
to polyvector extensions of Poincar\'e Lie superalgebras, but 
remark that these extensions have likewise been constructed 
and classified using the language of admissible bilinear forms 
\cite{Alekseevsky:2003vw}.

Starting from a bilinear form 
\[
\beta\;: {\bf S} \otimes {\bf S} \rightarrow \mathbb{R}
\]
on a spinor module ${\bf S}$, one can define a vector-valued bilinear form by
inserting a Clifford generator,
\[
K^\mu (\lambda, \chi) := 
\beta(\gamma^\mu \lambda\;, \chi) \;.
\]
One defines the symmetry $\sigma(\beta) = \pm 1$ and type
$\tau(\beta) = \pm 1$ of a bilinear form as follows:
\begin{align}
	\sigma(\beta) &: \beta(\lambda, \chi) = \sigma(\beta) \beta(\chi, \lambda) \;,\\
	\tau(\beta) &: \beta(\gamma^{\mu} \lambda, \chi) = \tau(\beta)\beta(\lambda, \gamma^{\mu}  \chi) \;.\nonumber
\end{align}
Note that the quantities $\sigma, \tau$ introduced earlier are the
symmetry $\sigma(C)$ and type $\tau(C)$ of the complex bilinear form
defined by the charge conjugation matrix $C$.

A  bilinear form is called {\em admissible} if it has a definite
symmetry and type. Note that this implies that it is
Spin$_{0}$-invariant,
since the generators of infinitesimal Lorentz transformations are
$\gamma^{\mu\nu}$. The Spin$_0$-equivariant vector-valued
bilinear form  associated to an admissible bilinear form $\beta$ is
symmetric if and only if $\sigma(\beta)\tau(\beta)=+1$.\footnote{Note
that so far we work with commuting spinors. Spinor fields with
anti-commuting components will be introduced later when we construct 
representations of the supersymmetry algebra on fields.}
We will call such vector-valued forms, and the underlying bilinear forms,
{\em super-admissible}.\footnote{This is a new terminology not used
in \cite{Alekseevsky:1997} which we find convenient.} 
Superbrackets, and the corresponding real, symmetric, Spin$_0$-equivariant
vector-valued bilinear forms form a vector space, and it was shown
in \cite{Alekseevsky:1997} that one can always choose a basis 
consisting of super-admissible vector-valued bilinear forms. 
Hence we can restrict ourselves to super-admissible bilinear forms in the
following. 

So far the spinor module ${\bf S}$ has been arbitrary. In the
following we focus on the case of the complex spinor module 
$\mathbb{S}$, that is on supersymmetry algebras where the 
supercharges form a single Dirac spinor. In odd dimensions,
where Dirac spinors are complex irreducible, this covers all the
minimal superalgebras, except in signatures where one can
obtain a smaller algebra by imposing a Majorana condition.
In even dimensions, where Dirac spinors decompose into
complex semi-spinors (Weyl spinors) one can also obtain
smaller supersymmetry algebras by imposing a chirality
condition. General extended superalgebras are obtained by
taking multiple copies of real irreducible spinor modules. 
We will leave these more complicated cases to future
work \cite{R-symmetry,Euc4d}, and focus on the case
where the supercharges form a single Dirac spinors. 
Moreover, when determining R-symmetry groups, we will
restrict to the case where Dirac spinors are complex irreducible, 
that is to odd dimensions.


\subsubsection{Standard Bilinear Forms}

The matrices $A$ and $C$ introduced previously naturally 
define admissible real bilinear forms on the complex spinor
module $\mathbb{S}$. 
Associated with the matrix $A$ we first 
have the Spin$_0$-invariant sesquilinear form
\[
A(\lambda, \chi) = \lambda^\dagger A \chi \;,\;\;\;\lambda, \chi \in
\mathbb{S} \;,
\]
which is either Hermitian or anti-Hermitian.
By decomposition of $A$ into its real and imaginary part we
obtain two admissible real bilinear forms, one symmetric, 
the other antisymmetric.

We can also define a  Spin$_0$-invariant complex bilinear form
\[
C(\lambda, \chi) = \lambda^T C \chi \;
\]
on $\mathbb{S}$ using a charge conjugation matrix $C$. This form 
has symmetry $\sigma(C)=\sigma$ and type
$\tau(C)=\tau$. By decomposition of $C$ 
into its real and imaginary part
we obtain two admissible real bilinear forms, which, depending on 
$\sigma$, are either both symmetric or both antisymmetric.

\subsubsection{R-symmetry groups}

The R-symmetry group of a supersymmetry algebra is the
group of all automorphisms of the superbracket
which commute with the Spin representation. Thus for a 
Poincar\'e Lie superalgebra $\mathfrak{sp}(V) = V + \mathfrak{so}(V) +
{\bf S}$ 
this is the group of all automorphisms which act trivially on the
even part $V+\mathfrak{so}(V)$. 
We first define the Schur algebra ${\cal C}({\bf S})$, 
which consists of all endomorphisms of ${\bf S}$ which are  
invariant under
the infinitesimal action of the Spin group:
\[
{\cal C}({\bf S}) = \{ Z \in \mbox{End}({\bf S}) | [Z,
\mathfrak{so}(V)] = 0  \} \;.
\]
The invertible elements of the Schur algebra form the Schur group
${\cal C}({\bf S})^*$. The R-symmetry group can then be defined as the 
subgroup of the Schur group which leaves the vector-valued bilinear
form $\beta(\gamma^\mu \cdot, \cdot)$ defining the superbracket invariant:
\[
G_R = \{ Z \in {\cal C}({\bf S})^* | \beta(\gamma^\mu Z \cdot, Z\cdot) 
= \beta(\gamma^\mu \cdot, \cdot) \} \;.
\]

\subsection{Doubled  Spinors}

We now introduce doubled spinors which will allow us to define
complex supersymmetry algebras in terms of complex vector-valued
bilinear forms. We take the spinor module ${\bf S}$ to be the complex 
spinor module $\mathbb{S}$, deferring the general case to a separate
publication \cite{R-symmetry}. We also assume that $\mathbb{S}$ is
complex irreducible, which is true in odd dimensions. We do not
impose that $\mathbb{S}$ is irreducible as a real representation, that is, in
signatures where Majorana spinors are possible one needs to analyze
separately whether one can define a smaller supersymmetry algebra. 
While we show that this does not happen in five dimensions, the 
general case is left to \cite{R-symmetry}.

Taking the spinor module to be $\mathbb{S}$, the 
doubled spinor module is $\mathbb{S} \oplus \mathbb{S}$. 
Since these are two copies of the same Spin$_0(t,s)$ representation
we can write this space as a complex tensor product between $\mathbb{S}$ and
an internal `multiplicity space' $\mathbb{C}^2$:
\[
\underline{\lambda} = 
(\lambda^i)_{i=1,2} = (\lambda^1, \lambda^2) \in \mathbb{S} \oplus 
\mathbb{S} \simeq \mathbb{S} \otimes \mathbb{C}^2 \ni (\lambda^i_\alpha) \;,
\]
where $\alpha = 1\, \ldots \dim \mathbb{S}$. This product form is
convenient since now the  Spin group only acts on the first factor, 
while, as we will show, the R-symmetry group only acts on the second factor.
Since $\mathbb{S}$ is a complex Spin$_0(t,s)$ module it
automatically carries a representation of the complex Spin group. 
By equipping $\mathbb{S}\otimes \mathbb{C}^2$ with a super-admissible
complex vector-valued bilinear form, we therefore obtain a complex
supersymmetry algebra.  To pick a real form for a given signature
$(t,s)$, we need to choose a Spin$_0$(t,s) invariant real 
structure $\rho$ on $\mathbb{S}\otimes \mathbb{C}^2$ and restrict 
the complex vector-valued bilinear form to the real points 
$(\mathbb{S} \otimes \mathbb{C}^2)^\rho$ with
respect to $\rho$. This way we obtain a super-admissible real vector-valued 
bilinear form on $(\mathbb{S}\otimes
\mathbb{C}^2)^\rho \simeq \mathbb{S}$. 
We now explain this construction in detail.

On any complex spinor module $\mathbb{S}$ there exists at least one
admissible complex bilinear form, the one defined by the charge
conjugation matrix $C$. This bilinear form can be used to define a
super-admissible complex vector-valued bilinear form on
$\mathbb{S}\otimes \mathbb{C}^2$. However,  since there might be 
other admissible  complex bilinear forms on $\mathbb{S}$, 
we use the generic notation
\[
\beta(\lambda, \chi) = \lambda^T Q_\beta \chi \;,\;\,\lambda, \chi \in \mathbb{C}
\]
in the following. Given any admissible bilinear form $\beta$ on $\mathbb{S}$, 
we obtain admissible complex bilinear forms on
$\mathbb{S} \otimes \mathbb{C}^2$ by tensoring $\beta$ with any 
non-degenerate symmetric
or antisymmetric complex bilinear form
\[
M\;: \mathbb{C}^2 \ni (w,z) \mapsto M(w,z) = w^i z^j M_{ji} 
\]
on $\mathbb{C}^2$. The resulting bilinear form $b=b_{\beta,M}=\beta\otimes M$,
\[
(\beta \otimes M)(\underline{\lambda}, \underline{\chi}) =
(\lambda^i)^T Q_\beta 
\chi^j M_{ji}
\]
has symmetry $\sigma(b_{\beta,M}) = \sigma(\beta) \sigma(M)$ and
type $\tau(b_{\beta,M})=\tau(\beta)$. 

To obtain a complex supersymmetry algebra
with associated spinor module 
$\mathbb{S}\otimes \mathbb{C}^2$,  we require that the 
complex vector-valued billinear form $b_{\beta,M}$ is
super-admissible, that is the associated vector-valued form must be 
symmetric:  $\sigma(b_{\beta,M}) \tau(b_{\beta,M})=+1$. Given a bilinear
form $\beta$ with $\sigma(\beta) \tau(\beta)=1$, one can therefore use any
symmetric bilinear form on $\mathbb{C}^2$, while for 
$\sigma(\beta) \tau(\beta)=-1$ we can use any antisymmetric bilinear form.

Since we want to determine the R-symmetry groups of the real
supersymmetry algebas later, it is 
useful to define complex versions of the Schur algebra
and of the R-symmetry group. The Schur
algebra ${\cal C}(\mathbb{S} \otimes \mathbb{C}^2)$ is defined as the subalgebra
of $\mbox{End}_{\mathbb{C}} (\mathbb{S}\otimes \mathbb{C}^2)$ which 
commutes with the Spin$_0$-representation. The invertible elements of 
the Schur algebra form the Schur group ${\cal C}^*(\mathbb{S}) 
\subset \mbox{Aut}_{\mathbb{C}} (\mathbb{S}\otimes \mathbb{C}^2)$.
Since we assume that $\mathbb{S}$ is
irreducible as a complex Spin$_0$-representation, we can apply Schur's 
lemma which implies that 
such endomorphisms
must act on $\mathbb{S}$ as a (non-zero) complex multiple of the unit 
$\mathbbm{1}$. This 
implies that 
${\cal C}^*(\mathbb{S} \otimes \mathbb{C}^2) \simeq GL(\mathbb{C}^2) \simeq
GL(2,\mathbb{C})$. We define the complex form of the R-symmetry group
to be the subgroup of ${\cal C}^*(\mathbb{S} \otimes \mathbb{C}^2)$
which leaves the complex vector-valued bilinear form $(\beta\otimes M)
(\gamma^\mu \cdot, \cdot)$ invariant. Since $M$ is either a symmetric
or an antisymmetric non-degenerate bilinear form, there are only two
cases:
\[
G_R^\mathbb{C} \simeq \left\{ \begin{array}{ll}
Sp(2,\mathbb{C})\;, \;\;\;&\mbox{if}\;M\;\mbox{is antisymmetric} \;,\\
O(2,\mathbb{C}) \;,\;\;\;& \mbox{if}\;M\;\mbox{is symmetric} \;.\\
\end{array} \right.
\]
The R-symmetry group of the real theory will then be a real form of
$G_R^\mathbb{C}$. For $G_R^{\mathbb{C}}=Sp(2,\mathbb{C})$ the real 
R-symmetry group is locally isomorphic to 
$SU(2)$ or to $SU(1,1)$, while
for $G_R^{\mathbb{C}}=O(2,\mathbb{C})$ it is locally isomorphic to  
$SO(2)$ or to $SO(1,1)$. 

Note that in order to apply Schur's lemma, we need to assume that
$\mathbb{S}$ is complex irreducible. 
If $\mathbb{S}$ is complex reducible, as it happens in even dimensions, 
then the R-symmetry group will be extended by transformations that
act non-trivially on the factor $\mathbb{S}$ of $\mathbb{S} \otimes 
\mathbb{C}^2$. See for example \cite{Cortes:2003zd}.

\subsection{Reality Conditions}

We now show how to recover Dirac spinors from  doubled spinors 
by imposing 
a Spin$_0$-invariant reality condition.
The possible reality conditions depend on whether $\mathbb{S}$
carries a real or quaternionic structure. 

\subsubsection{Symplectic Majorana spinors}

We start with the case $BB^*=-\dblone$, where $\mathbb{S}$ carries
a Spin$_0$-invariant quaternionic structure
$j^{(\alpha)}_{\mathbb{S}}$. To obtain a Spin$_0$-invariant
real structure on $\mathbb{S}\otimes \mathbb{C}^2$, we can take the
product with any quaternionic structure $j_{\mathbb{C}^2}$ on
  $\mathbb{C}^2$.  For concreteness, we take the standard quaternionic
structure on $\mathbb{C}^2$, 
\[
j_{\mathbb{C}^2} \left[ \left( \begin{array}{c}
z_1 \\ z_2 \\
\end{array} \right)\right] = \left( \begin{array}{c}
- z_2^* \\ z_1^* \\
\end{array} \right) \;.
\]
The quaternionic structure on $\mathbb{S}$ is
\[
j_{\mathbb{S}} (\lambda) = \alpha^* B^* \lambda^* \;.
\]
The resulting real structure $\rho=j_{\mathbb{S}} \otimes
j_{\mathbb{C}^2}$ is
\[
\rho\left[\left( \begin{array}{cc}
\lambda^1 \\ \lambda^2 \\
\end{array} \right)\right] = 
\left( \begin{array}{c} 
- j_{\mathbb{S}} \lambda^2 \\
j_{\mathbb{S}} \lambda^1 \\
\end{array} \right) =
\left( \begin{array}{cc}
\alpha^* B^* (\lambda^{i})^* \varepsilon_{i1} \\
 \alpha^* B^* (\lambda^{i})^* \varepsilon_{i2} \\
\end{array} \right) \;,
\]
where
\[
\varepsilon = \begin{pmatrix}
		0 & 1 \\ -1 & 0
	\end{pmatrix} \;.
\]
Symplectic Majorana spinors are by definition those doubled spinor
which are real in the sense that they are invariant under $\rho$:
\[
\rho(\lambda)= \lambda \Leftrightarrow \lambda^2 = j_{\mathbb{S}}
\lambda^1 \Leftrightarrow (\lambda^i)^* = \alpha B \lambda^j
\varepsilon_{ji} \;.
\]
Since we can identify the subspace $\mathbb{S}_{SM} =
(\mathbb{S}\otimes \mathbb{C}^2)^\rho 
\subset \mathbb{S} \otimes
\mathbb{C}^2 \simeq \mathbb{S} \oplus \mathbb{S}$ of real doubled
spinors with the graph
of $j_{\mathbb{S}}$ on $\mathbb{S}$, we obtain an isomorphism between
Dirac spinors and symplectic Majorana spinors
\[
\mathbb{S} \ni \lambda^1 \mapsto (\lambda^1,
j_{\mathbb{S}}(\lambda^1)) \in \mathbb{S}_{SM} \subset
\mathbb{S}\otimes \mathbb{C}^2 \;.
\]
This is an isomorphism of real Clifford modules and hence of real Spin
modules \cite{Cortes:2003zd}. The elements $\phi \in GL(2,\mathbb{C})
\simeq {\cal C}^*(\mathbb{S}\otimes \mathbb{C}^2)$ of the
Schur group which
commute with the reality condition form a subgroup. 
Evaluating the condition 
\[
j_{\mathbb{C}^2} \varphi (\lambda) = \varphi j_{\mathbb{C}^2} (\lambda)
\;,\;\mbox{where}\;\;
j_{\mathbb{C}^2} (\lambda^i) = (\lambda^j)^* \varepsilon_{ji} \;.
\]
we find
\begin{equation}
\label{SU2}
\varphi = \left( \begin{array}{cc}
u & v \\
-{v}^* & {u}^* \\
\end{array} \right) \in GL(1,\mathbb{H}) \subset GL(2,\mathbb{C}) \;,
\end{equation}
where $GL(1,\mathbb{H})=\mathbb{H}^*$ is the group of invertible
quaternions. The corresponding R-symmetry will be worked out below.

\subsubsection{Real doubled spinors defined by a Majorana condition}

In signatures where $B^*B=\dblone$, one can define Majorana spinors
and possibly obtain a supersymmetry algebra that is smaller than the
one based on Dirac spinors. 
But we can also use the Majorana condition to 
to describe Dirac spinors in terms of real doubled 
spinors. This time the reality condition imposed on doubled
spinors is the product of the Spin$_0$-invariant
real structure $\rho_{\mathbb{S}}$ on $\mathbb{S}$ defined by the
Majorana condition, and a real
structure on $\mathbb{C}^2$. For later convenience we choose 
\begin{equation}
\label{Real_structure_doubled}
\rho_{\mathbb{C}^2}  \left[ \left( \begin{array}{c}
z_1 \\ z_2 \\
\end{array} \right)\right] = \left( \begin{array}{c}
z_2^* \\ z_1^* \\
\end{array} \right) \;.
\end{equation}
The real structure on $\mathbb{S}$ is
\[
\rho_{\mathbb{S}}(\lambda) = \alpha^* B^* \lambda^* \;.
\]
The resulting real structure on $\mathbb{S}\otimes \mathbb{C}^2$ is
\[
\rho\left[\left( \begin{array}{c}
\lambda^1 \\ \lambda^2 \\
\end{array} \right)\right] = 
\left( \begin{array}{c} 
 \rho_{\mathbb{S}} \lambda^2 \\
\rho_{\mathbb{S}} \lambda^1 \\
\end{array} \right) =
\left( \begin{array}{c}
\alpha^* B^* (\lambda^{i})^* \eta_{i1} \\ 
\alpha^* B^* (\lambda^{i})^* \eta_{i2} \\
\end{array} \right) \;,
\]
where 
\[
(\eta_{ij}) = \left( \begin{array}{cc}
0& 1\\
1 & 0 \\
\end{array} \right) \;.
\]
Real doubled spinors 
are those doubled spinors 
which are invariant under $\rho$:
\[
\rho(\lambda)= \lambda \Leftrightarrow \lambda^2 = \rho_{\mathbb{S}}
\lambda^1 \Leftrightarrow (\lambda^i)^* = \alpha B \lambda^j
\eta_{ji} \;.
\]
Similarly to the symplectic Majorana case, we have an isomorphism
between Dirac spinors and real doubled spinors using that 
the real doubled spinors lie on the graph of the real structure:
\[
\mathbb{S} \ni \lambda^1 \mapsto (\lambda^1,
\rho_{\mathbb{S}}(\lambda^1)) \in \mathbb{S}_{DM} \subset
\mathbb{S}\otimes
\mathbb{C}^2 \simeq \mathbb{S} \oplus \mathbb{S} \;.
\]
To find the Schur group associated with the real theory
we need to identify those elements $\varphi\in GL(2,\mathbb{C})\simeq
{\cal C}^*(\mathbb{S}\otimes \mathbb{C}^2)$ 
which commute 
with the real 
structure $\rho_{\mathbb{C}^2}$. The condition is
\[
\rho_{\mathbb{C}^2} \varphi (\lambda) = \varphi \rho_{\mathbb{C}^2} (\lambda)
\;,\;\mbox{where}\;\;
\rho_{\mathbb{C}^2} (\lambda^i) = (\lambda^j)^* \eta_{ji} \;.
\]
which implies
\begin{equation}
\label{SU11}
\varphi = \left( \begin{array}{cc}
u & v \\
{v}^* & {u}^* \\
\end{array} \right) \in GL(1,\mathbb{H}') \subset GL(2,\mathbb{C}) \;,
\end{equation}
where $GL(1,\mathbb{H}')\simeq GL(2,\mathbb{R}) \subset
GL(2,\mathbb{C})$ 
is the group of invertible
para-quaternions.\footnote{See Appendix \ref{App-para-quaternions} for
  the relevant fact about para-quaternions.}

\subsection{Determination of R-symmetry groups}

We have seen that the R-symmetry group of the complex theory is
either isomorphic to $O(2,\mathbb{C})$ or to $Sp(2,\mathbb{C})=SL(2,\mathbb{C})$,
depending on whether the complex bilinear form we choose on
$\mathbb{C}^2$ is symmetric or antisymmetric. The possible real R-symmetry
groups are those subgroups which commute with the reality condition we
impose. Therefore the real R-symmetry groups are real forms of the
complex R-symmetry groups. For semi-simple groups, such as
$Sp(2,\mathbb{C})$, this implies that they can be obtained as fixed
points under an involutive automorphism, see for example \cite{Gilmore} for 
background. The group $O(2,\mathbb{C})$ 
is abelian, and not semi-simple.  Since there are only
two connected real one-dimensional Lie groups, $SO(2)$ and
$\mathbb{R}\simeq SO_0(1,1)$, we know that these are the
connected components of the real R-symmetry groups. We will show
that both the compact and the non-compact form appear, and 
determine the group globally for the standard involutions.

Let us show explicitly how imposing a reality condition on doubled
spinors is related to involutive automorphisms of $G_R^{\mathbb{C}}$. 
The real R-symmetry group is defined by
\[
G_R = \{ \varphi \in G_R^\mathbb{C} | \rho \varphi = \varphi \rho \} 
\subset G_R^\mathbb{C} \simeq GL(2, \mathbb{C}) \;,
\]
where $\rho$ is either a real structure $\rho_{\mathbb{C}^2}$ or a 
quaternionic structure $j_{\mathbb{C}^2}$ on $\mathbb{C}^2$. Its
action takes the form
\[
\rho(z^i) = z^{j*} N_{ji} \;,\;\;\;i,j=1,2 \;,
\]
where $N=(N_{ji})$ is a real invertible matrix which is either
symmetric or antisymmetric, and satisfies $N^2 = \pm \mathbbm{1}$. 
The choices corresponding
to the two types of real doubled spinors introduced earlier
are $\eta_{ji}$ and $\varepsilon_{ji}$. We parametrize the action of a complex
R-symmetry transformation  $\varphi \in G_R^\mathbb{C}$ as
\[
\varphi(z^i) = A^i_{\;\;j} z^j\;,
\]
where $A=(A^i_{\;\;j}) \in GL(2,\mathbb{C})$. Evaluating the condition 
$\rho \varphi = \varphi \rho$ we find, taking into account that
$N^T=\pm N$, 
\[
A = N A^* N^{-1}
\]
This is precisely the action of an involutive automorphism acting by
complex conjugation $K$ composed with conjugation by the matrix $N$:
\[
A \mapsto N \circ K (A) = N K(A) N^{-1} \;.
\]
Since we would like to know the R-symmetry groups for all possible
types of reality conditions, we first need to review the relevant facts about
real forms of complex Lie groups. For a connected 
simple complex Lie group, all real forms can be obtained as fixed
points of involutive automorphisms.  Up to conjugation with 
inner automorphisms, there are at most three types of such automorphisms:
complex conjugation $K$, and the conjugation by matrices of the form
\[
I_{p,q} = \left( \begin{array}{cc}
\mathbbm{1}_p & 0 \\
0 & - \mathbbm{1}_q \\
\end{array} \right) \;,\;\;\;
J_{p,p} = \left( \begin{array}{cc}
0 & \mathbbm{1}_p \\
- \mathbbm{1}_q & 0\\ 
\end{array} \right) \;.
\]
For the connected simple group
$G_R^{\mathbb{C}} = Sp(2,\mathbb{C}) = SL(2,\mathbb{C})$, the three
basic involutions are $K$, $I_{1,1}$ and $J_{1,1} = \varepsilon$. 
We find it convenient to replace $I_{1,1}$ by the equivalent matrix $\eta$,
and to use $K, \eta \circ K, \varepsilon \circ K$ as the independent
involutions. The group $G_R^{\mathbb{C}} = O(2,\mathbb{C})$ has two
connected components and is abelian rather than simple. Since there
are only two connected
one-dimensional Lie groups, $SO(2)\simeq U(1)$ and $SO_0(1,1)\simeq
\mathbb{R}$ the subgroup of $O(2,\mathbb{C})$ consistent with the
reality condition must be locally isomorphic to one of them. 

To determine the real R-symmetry groups, we need to solve the equation 
\begin{equation}
\label{A}
A^* = N^{-1} A N
\end{equation}
for $A\in Sp(2,\mathbb{C}), O(2,\mathbb{C})$ and $N=\mathbbm{1}, \eta,
\varepsilon$. Working infinitesimally, $A=\exp(\epsilon a)=
\mathbbm{1} + \epsilon a + O(\epsilon^2)$, we see that the Lie algebra element $a$
must satisfy the same equation as the group element, 
\begin{equation}
\label{a}
a^* = N^{-1} a N \;.
\end{equation}
We start with $A\in Sp(2,\mathbb{C})$, where
\[
a = \left( \begin{array}{cc}
\alpha & \beta \\
\gamma & - \alpha \\
\end{array} \right) \;,\;\;\;\alpha, \beta, \gamma \in \mathbb{C} \;.
\]
\begin{itemize}
\item $N=\mathbbm{1}$. \\
In this case $a$ and $A$ must be real, which implies $G_{R} =
Sp(2,\mathbb{R})
\simeq SU(1,1)$.
\item $N=\eta$.\\
Equation (\ref{a}) implies $\alpha^* = -\alpha$, $\beta = \gamma^*$. 
Writing $\alpha = \alpha_1 + i \alpha_2$, etc. we obtain
\[
a = \left( \begin{array}{cc}
i\alpha_2 & \beta_1 + i \beta_2 \\
\beta_1 - i \beta_2 & - i\alpha_2 \\
\end{array} \right) \;.
\]
The Lie algebra spanned by these matrices has two non-compact
generators, and therefore is  isomorphic to $\mathfrak{su}(1,1)$. By 
exponentiation we obtain $G_R \simeq SU(1,1)$. 
\item $N=\varepsilon$.\\
Equation (\ref{a}) implies $\alpha^* = -\alpha$, $\beta = -\gamma^*$. 
Therefore
\[
a = \left( \begin{array}{cc}
i\alpha_2 & \beta_1 + i \beta_2 \\
- \beta_1 + i \beta_2 & - i\alpha_2 \\
\end{array} \right) \;.
\]
The Lie algebra spanned by these matrices has three compact
generators, and is therefore isomorphic to $\mathfrak{su}(2)$. 
By exponentiation we obtain $G_R\simeq SU(2)$.
\end{itemize}
We remark that the R-symmetry groups $SU(2)$ and $SU(1,1)$ 
obtained for $N=\varepsilon, \eta$ have a natural interpretation 
in terms of the quaternionic and para-quaternionic structure on the
complex spinor module $\mathbb{S}$. For $N=\varepsilon$ the real
Schur group was found to be $GL(1,\mathbb{H}) \subset
GL(2,\mathbb{C})$ and we can obtain the real R-symmetry group by
intersecting this with the complex R-symmetry group. This reveals that 
the R-symmetry group
\[
GL(1,\mathbb{H}) \cap Sp(2,\mathbb{C}) = U(1,\mathbb{H}) = SU(2) \;,
\]
is the group of unit norm quaternions, which rotates the
Spin$_0$-invariant quaternionic structures $I,J,K$ on $\mathbb{S}$.

Similarly, for $N=\eta$ the real Schur group was found to be
$GL(1,\mathbb{H}') \subset GL(2,\mathbb{C})$.
The real R-symmetry group obtained by intersecting with the complex
R-symmetry group,
\[
GL(1,\mathbb{H}') \cap Sp(2,\mathbb{C}) = U(1,\mathbb{H}')=SU(1,1)  \;,
\]
is the group of unit norm para-quaternions, which 
rotates the Spin-invariant 
para-quaternionic structure $I,J,K$ on $\mathbb{S}$.


We now turn to $G_R^{\mathbb{C}} = O(2,\mathbb{C})$. An element
of the complex Lie algebra $\mathfrak{o}(2,\mathbb{C})$ is an
antisymmetric
complex two-by-two matrix
\[
\left( \begin{array}{cc}
0  & \alpha  \\
-\alpha  & 0 \\
\end{array} \right) \;,\;\;\;\alpha \in \mathbb{C} \;.
\]
\begin{itemize}
\item $N=\mathbbm{1}$. 
In this case $a$ and $A$ must be real and we obtain 
$G_R=O(2)$. 
\item $N=\eta$. \\
In this case (\ref{a}) implies $\alpha^* = - \alpha$, so that
\[
a = \left( \begin{array}{cc} 
0 & i \alpha_2 \\
- i \alpha_2 & 0\\
\end{array} \right) \;,\;\;\;\alpha_2 \in \mathbb{R}\;.
\]
This generates a non-compact group, therefore $G_R \supset SO(1,1)_0
\simeq \mathbb{R}$. 

Since $G_R^{\mathbb{C}}= O(2,\mathbb{C})$ has two connected
components, while the largest possible real R-symmetry group $O(1,1)$ 
has four connected components, 
we need additional work to determine the group
globally. We use the parametrization
\[
\left( \begin{array}{cc}
\cos z & \sin z \\
- \sin z & \cos z \\
\end{array} \right) \in SO(2,\mathbb{C}) \;,\;\;\;
\left( \begin{array}{cc}
\cos z & \sin z \\
\sin z & - \cos z \\
\end{array} \right) \in O(2,\mathbb{C}) \backslash SO(2,\mathbb{C}) \;.
\]
For $A \in SO(2,\mathbb{C})$ equation (\ref{A}) implies
\[
(\cos z)^* = \cos z \;,\;\;\;\;
(\sin z)^* = - \sin z  \;.
\]
This has two solutions for $z$, 
\[
z = i\chi \;,\;\;\;z = \pi + i \chi \;,\;\;\;\chi\in \mathbb{R} \;.
\]
The resulting matrices $A$ take the form
\[
A = \pm \left( \begin{array}{cc}
\cosh \chi & i \sinh \chi \\
- i \sinh \chi & \cosh \chi \\
\end{array} \right)\simeq 
\pm \left( \begin{array}{cc}
\cosh \chi & \sinh \chi \\
 \sinh \chi & \cosh \chi \\
\end{array} \right) \;,\;\;\;\chi\in \mathbb{R} \;,
\]
where we made an equivalence transformation corresponding to the map
$(z^1, z^2) \mapsto (z^1, i z^2)$ in the second step. After this
transformation $A$ takes the standard form of an $SO(1,1)$ matrix, and
we see that the group we obtain by imposing the reality condition on
$SO(2,\mathbb{C})$ is isomorphic to $SO(1,1)$. 

For $A \in O(2,\mathbb{C}) \backslash SO(2,\mathbb{C})$, equation
(\ref{A}) implies
\[
(\cos z)^* = -\cos z \;,\;\;\;\;
(\sin z)^* =  \sin z  \;.
\]
This again has two solutions for $z$:
\[
z = \frac{\pi}{2} + i \chi \;,\;\;\;
z = \frac{3\pi}{2} + i \chi \;,\;\;\;
\chi \in \mathbb{R} \;.
\]
The corresponding matrices $A$ take the form
\[
A = \pm \left( \begin{array}{cc}
- i \sinh \chi & \cosh \chi \\
\cosh \chi & i \sinh \chi \\
\end{array} \right)  
\]
These matrices have determinant $-1$ and extend the R-symmetry group 
to $O(1,1)$. 
\item
$N=\varepsilon$. \\
At the infinitesmal level equation (\ref{a}) implies $\alpha =
\alpha^*$, so that
\[
a = \left( \begin{array}{cc}
0 & \alpha_1 \\
- \alpha_1 & 0 \\
\end{array} \right) \;,\;\;\;\alpha_1 \in \mathbb{R} \;.
\]
This generates a compact group, so $G_R \supset SO(2)$. To decide 
whether the group is $O(2)$ or $SO(2)$, we turn to equation (\ref{A}).
For $A\in SO(2,\mathbb{C})$ we obtain:
\[
(\cos z)^* = \cos(z) \;,\;\;\;
(\sin z)^* = \sin(z) 
\]
so that
\[
A = \left( \begin{array}{cc}
\cos \phi & \sin \phi \\
-\sin \phi & \cos \phi \\
\end{array} \right)
\]
as it must, given the result for the Lie algebra. For $A\in
O(2,\mathbb{C}) \backslash SO(2,\mathbb{C})$ we obtain
\[
(\cos z)^* = -\cos z \;,\;\;\;
(\sin z)^* = - \sin z  \;.
\]
These two equations have no common solution, therefore the
R-symmetry group is $G_R = SO(2)$. 
\end{itemize}
We note that the differences between the R-symmetry groups
$O(2), SO(2), O(1,1)$ can be characterised in terms of their action 
on $\mathbb{R}^2 \subset \mathbb{C}^2$. While on $\mathbb{C}^2$ 
the signature of a complex bilinear form is not an invariant, the
signature of a real bilinear form on $\mathbb{R}^2$ is an invariant, 
which distinguishes $O(2)$ and $O(1,1)$. Moreover, on $\mathbb{R}^2$
we can choose an orientation, or a complex structure, which is
preserved by $SO(2) \subset O(2)$, while the full group $O(2)$ also
contains transformations which reverse orientation and complex
structure. Similarly, $\mathbb{R}^2$ can be equipped by a para-complex
structure, which is preserved by $SO(1,1)\subset O(1,1)$ only.

This completes the determination of the R-symmetry groups that can
appear in supersymmetry algebras based a single Dirac spinor of
supercharges in odd dimensions. 
We now specialize to five dimensions where we will show explicitly how
to relate superbrackets on $\mathbb{S}$ and $\mathbb{S} \otimes
\mathbb{C}^2$ to one another.

{\section{Minimal Supersymmetry in Five Dimensions \label{Sect:5dsusy}}

We first collect a few useful facts and relations valid in five
dimensions. The complex spinor module $\mathbb{S}$ 
is irreducible as a complex module. We will see that for some
signatures Majorana spinors exist, which means that $\mathbb{S}$
is reducible as a real module, and that smaller, ${\cal N}=1$
superalgebras might exist. However, as we show at the end of this section
these superalgebras are trivial in the sense that the supercharges
simply anti-commute. 
From the tables in \cite{Alekseevsky:1997} we can read off that
there the vector space of symmetric Spin$_0$-equivariant vector-valued
bilinear forms is one-dimensional. In other words, for all signatures
the superbracket is unique up to rescaling. We
will provide explicit realizations of the superbrackets 
both in terms of Dirac spinors and in terms of doubled Dirac spinors.

The charge
conjugation matrix $C$ is unique up to equivalence and has symmetry
$\sigma=\sigma(C)=-1$ and type $\tau=\tau(C)=1$. It is therefore 
antisymmetric, $C^T=-C$. Since $C$ satisfies $C^{-1}=C=C^\dagger$, it
is then purely imaginary, $C^*=-C$. 
Since $\tau=1$,
\[
(\gamma^\mu)^T = \tau C \gamma^\mu C^{-1} = C \gamma^\mu C^{-1}\;.
\]
A straightforward calculation shows that 
$B=(CA^{-1})^T$ can be re-written as
\begin{equation}
\label{B=CA}
B = (-1)^{t+1} CA \;,
\end{equation}
and complex conjugation operates on $\gamma$-matrices by
\[
(\gamma^\mu)^* = (-1)^t \tau B\gamma^\mu B^{-1} =
(-1)^t  B\gamma^\mu B^{-1} \;.
\]
By another straightforward calculation we find
\[
B^* B = (-1)^{(t^2 + 3t +2)/2} \dblone = \left\{
\begin{array}{ll} 
- \dblone\;, & \mbox{for}\;t=0,1,4,5\;, \\
+ \dblone\;, & \mbox{for}\;t=2,3 \;.
\end{array} \right. 
\]
Thus the spinor module $\mathbb{S}$ has an invariant quaternionic
structure for 
Euclidean and Lorentz signature, but an invariant real structure for signatures
with two times. For later use we note that
\begin{equation}
\label{BC=A}
B^\dagger C = (-1)^{(t^2+3t+2)/2} A = \left\{ \begin{array}{ll}
-A & \mbox{for}\;t=0,1,4,5 \\
A & \mbox{for}\; t=2, 3 \;. \\
\end{array} \right. 
\end{equation}

To define the five-dimensional $\mathcal{N}=2$ superalgebras 
in terms of Dirac spinors, we use the sesquilinear form
\[
A(\lambda, \chi) = \lambda^\dagger A \chi \;.
\]
$A$ is Hermitian for $t=0,3,4$ and anti-Hermitian for
$t=1,2,5$. The real and imaginary part of $A$ define 
two admissible real bilinear forms, whose symmetry 
is determined by the Hermiticity properties of $A$. The
type of $A$, $\mbox{Re}(A)$ and $\mbox{Im}(A)$ is $+1$ if
$A$ is a product of an even number of generators and $-1$ if
$A$ is a product of an odd number of generators. The resulting
values for the invariants $\sigma$ and $\tau$ are summarized in 
Table \ref{5dbilinearforms}. 
\begin{table}[h!]
\centering
\begin{tabular}{|l|ll|} \hline
 & $\mbox{Re}[A]$ & $\mbox{Im}[A]$  \\ 
 \hline $(0,5)$ & $(+,+)$ & $(-, +)$  \\ 
 \hline $(1,4)$ & $(-,-)$ & $(+, -)$  \\ 
 \hline $(2,3)$ & $(-,+)$ & $(+, +)$  \\  
 \hline $(3,2)$ & $(+,-)$ & $(-, -)$  \\ 
 \hline $(4,1)$ & $(+,+)$ & $(-, +)$ \\ 
 \hline $(5,0)$ & $(-,-)$ & $(+, -)$ \\ \hline
\end{tabular}
\caption{Invariants, $(\sigma = \pm 1, \tau = \pm 1)$, of Dirac
  bilinear forms in various signatures \label{5dbilinearforms}} 
\end{table}
The superbrackets corresponding to the vector-valued bilinear forms
$b_A(\gamma^\mu \cdot, \cdot) = \mbox{Re/Im} A(\gamma^\mu \cdot ,
\cdot)$ are defined by 
\[
\{ \lambda^\alpha Q_\alpha \;, \chi^\beta Q_\beta \} =
b_A(\gamma^\mu \lambda, \chi) P_\mu \;.
\]
 Using our spinor conventions summarized in Appendix \ref{Spinor_Con},
the corresponding anti-commutation relations are:
\begin{align}
	\{Q_{\alpha}, Q_{\beta} \} = \begin{cases} \mbox{Re}[\gamma^{\mu} A^{-1}]_{\alpha \beta} P_{\mu}\;, \quad t=0,1,4,5  \\
 \mbox{Im}[\gamma^{\mu} A^{-1}]_{\alpha \beta} P_{\mu}\;, \quad t=2,3\;.
 \end{cases}
\end{align}

We can also define an admissible complex bilinear form based on the
charge conjugation matrix $C$, 
\begin{align}
	C(\lambda, \chi) = \lambda^T C \chi \;,
\end{align}
which gives rise to two admissible 
real bilinear forms, $\mbox{Re}(C)$ and $\mbox{Im}(C)$.
Since the charge conjugation matrix is unique
(up to equivalence) with $\sigma=-1$, $\tau=1$, 
these forms are not superadmissible, see Table \ref{5dbilinearforms2}. 
\begin{table}[h!]
\centering
\begin{tabular}{|l|ll|} \hline
 & $Re[C]$ & $Im[C]$  \\ 
 \hline All signatures & $(-,+)$ & $(-, +)$  \\ \hline
 \end{tabular}
 \caption{Invariants, $(\sigma = \pm 1, \tau = \pm 1)$, of the charge
   conjugation bilinear forms in all five dimensional signatures \label{5dbilinearforms2}}
 \end{table}
However, we can 
define a complex superbracket on the doubled spinor module $\mathbb{S}\otimes 
\mathbb{C}^2$ using the admissible bilinear form
$b_{C,\varepsilon}= C \otimes \varepsilon$:
\[
b_{C,\varepsilon}({\lambda}, {\chi}) = 
(C\otimes \varepsilon)({\lambda}, {\chi}) =
{\lambda}^T C \chi =
(\lambda^i)^T C \chi^j \varepsilon_{ji} \;,
\]
and the corresponding vector-valued bracket $b_{C,
  \varepsilon}(\gamma^\mu {\lambda}, {\chi})$.

\subsection{Real doubled spinors in $(0,5), (1,4), (4,1)$ and $(5,0)$
  - Using the symplectic
  Majorana condition}

In signatures with no or one time-like dimension, 
the superbracket on $\mathbb{S}$ is given by the real part of
\[
A(\gamma^\mu \lambda^1, \chi^1) =
(-1)^t (\lambda^1)^\dagger A \gamma^\mu \chi^1 \;,
\]
where $\lambda^1, \chi^1\in \mathbb{S}$. 
Since $B^*B=-\dblone$, the complex spinor module $\mathbb{S}$ carries
a 
Spin$_0$-invariant
quaternionic structure $j_{\mathbb{C}^2}$. Therefore we can define
symplectic Majorana spinors using the real structure
$\rho=j_{\mathbb{S}} \otimes j_{\mathbb{C}^2}$. To evaluate the 
super-admissible bilinear form $b_{C,\varepsilon}(\gamma^\mu \cdot,
\cdot)$ on the space $\mathbb{S}_{SM} = (\mathbb{S} \otimes
\mathbb{C}^2)^\rho$ of real doubled spinors, we use the reality condition to
eliminate the second components of doubled spinors, 
\[
\lambda^2 = \alpha^* B^* (\lambda^1)^* \;.
\]
A short calculation shows\footnote{From now on we will stop 
`underlining' doubled spinors, as it should be clear from context 
whether $\lambda$ refers to a Dirac spinor or a doubled spinor.}
\begin{eqnarray}
b_{C,\varepsilon} (\gamma^\mu \lambda, \chi)_{\rho} 
&=& 
[(\gamma^\mu \lambda^2)^T C \chi^1 - (\gamma^\mu \lambda^1)^T
C \chi^2]_{\rho} \nonumber \\
&=& \alpha^* \left[  (\lambda^1)^\dagger B^\dagger (\gamma^\mu)^T C \chi^1
- (\lambda^1)^T (\gamma^\mu)^T C B^* (\chi^1)^*  \right]\nonumber \\
&=& \alpha^* \left[ (-1)^{(t^2+3+2)/2} (\lambda^1)^\dagger A \gamma^\mu
    \chi^1
- ((\lambda^1)^\dagger A \gamma^\mu \chi^1)^* \right] \\
&=& (-1)^t \alpha^* \left( (-1)^{(t^2+3t+2)/2} ( (A\lambda^1)^\dagger
    \gamma^\mu \chi^1) - ( (A\lambda^1)^\dagger \gamma^\mu \chi^1)^*
    \right) \;. \nonumber
\end{eqnarray}
Here we used (\ref{B=CA}) and (\ref{BC=A}). 
Since $(-1)^{(t^2+3t+2)/2}=-1$ for $t=0,1,4,5$, we find
\[
(C \otimes \varepsilon)_{\rm real} (\gamma^\mu \lambda, \chi) =
-2 (-1)^t \alpha^* \mbox{Re}(A)(\gamma^\mu \lambda^1,\chi^1) \;.
\]
Up to a complex overall factor, the super-admissible 
vector-valued real bilinear 
forms on real doubled spinors and on Dirac spinors
agree and define
the same superalgebra. 

We make the conventional choice to rescale the complex bilinear form
on $\mathbb{S}\otimes \mathbb{C}^2$ by a factor $-\frac{1}{2}$ and
to set $\alpha = (-1)^t$. Then the restriction of $b(\gamma^\mu \cdot,
\cdot)=-\frac{1}{2}
(C\otimes \varepsilon)(\gamma^\mu \cdot, \cdot)$ to symplectic Majorana spinors agrees with
$\mbox{Re}(A)(\gamma^\mu \cdot, \cdot)$. Using the conventions summarized in Appendix
\ref{Spinor_Con}, the five-dimensional ${\cal N}=2$ superalgebra
takes the form
\[
\{ Q_{i \alpha} , Q_{i\beta} \} = - \frac{1}{2} (\gamma^\mu
C^{-1})_{\alpha \beta} P_\mu \varepsilon_{ij} \;,
\]
where the doubled spinors $Q_i=(Q_{i\alpha})$ are subject to the reality condition
\[
Q_i^* = (-1)^{t+1} B \varepsilon^{ij} Q_j
\Leftrightarrow
Q^{i*} = (-1)^t B Q^j \varepsilon_{ji} \;.
\]
Since $\mathbb{S}$ is irreducible and the bilinear
form $\varepsilon$ on $\mathbb{C}^2$ is antisymmetric, the R-symmetry  group is
\[
G_R = GL(1, \mathbb{H}) \cap SL(2, \mathbb{C}) \simeq SU(2) \;.
\]

\subsection{Real doubled spinors in $(2,3)$ and $(3,2)$ - Using the 
Majorana condition}

In signatures with two time-like dimensions
the superbracket on $\mathbb{S}$ is given by the imaginary part of
\[
A(\gamma^\mu \lambda^1, \chi^1) =
(-1)^t (\lambda^1)^\dagger A \gamma^\mu \chi^1 \;.
\]
In these signature $B^*B=\dblone$, so that $B$ defines a
Spin$_0$-invariant real structure $\rho_{\mathbb{S}}$ on $\mathbb{S}$,
which allows us to define
Majorana spinors. We show at the end of this section
that the
restriction of the superbracket to Majorana spinors is trivial and
does not define a smaller, ${\cal N}=1$ superalgebra with four real
supercharges. Here we focus on rewriting Dirac spinors as real doubled
spinors.  By combining the 
real structure $\rho_{\mathbb{S}}$ with the real structure
$\rho_{\mathbb{C}^2}$ on $\mathbb{C}^2$ defined by (\ref{Real_structure_doubled})
we obtain the Spin$_0$-invariant real structure
$\rho=\rho_{\mathbb{S}}\otimes \rho_{\mathbb{C}^2}$ on doubled spinors.
Evaluating the invariant bilinear form $b_{C,\varepsilon}(\gamma^\mu
\cdot, \cdot)$ on real doubled spinors, we find
\begin{eqnarray*}
b_{C,\varepsilon}(\gamma^\mu \lambda, \chi)_{|\rm real} &=& 
\alpha^* (\lambda^1)^\dagger B^\dagger C \gamma^\mu \chi^1 +
\alpha^* (-1)^t ( (\lambda^1)^\dagger CB \gamma^\mu \chi^1)^* \\
&=& \alpha^* (-1)^{t(t+1)/2} (A(\gamma^\mu\lambda^1, \chi^1) -
    A(\gamma^\mu \lambda^1,\chi^1)^*)  \\
&=& \alpha^* (-1)^{t(t+1)/2} 2i \mbox{Im}(A) (\gamma^\mu \lambda^1,\chi^1) \;,
\end{eqnarray*}
where we used (\ref{B=CA}), (\ref{BC=A}) and 
that $(-1)^{(t^2+3t+2)/2} = 1$ for $t=2,3$. 
As for the other signatures, we rescale the bilinear on
$\mathbb{S}\otimes \mathbb{C}^2$ by a factor $-\frac{1}{2}$,
so that 
\[
b(\gamma^\mu \lambda,\chi)_{|\rm real} = - i (-1)^{t(t+1)/2}\alpha^*
\mbox{Im}(A)(\gamma^\mu \lambda^1,\chi^1) \;.
\]
Then by choosing
\[
\alpha = -i (-1)^{t(t+1)/2} = \left\{ \begin{array}{ll} 
i & \mbox{for}\;t=2\;,\\
-i & \mbox{for}\;t=3\;,\\
\end{array}
\right.
\]
the invariant vector-valued  bilinear form on $\mathbb{S}\otimes
\mathbb{C}^2$ agrees with the invariant vector-valued real bilinear 
form on $\mathbb{S}$ when evaluated on real points. 
Note that with
our conventional choices of $\alpha$, the five-dimensional
supersymmetry algebra takes the same form
for all signatures when expressed using doubled spinors. The reality 
conditions which relate doubled spinors to Dirac spinors are
summarized
in Table \ref{Values_alpha}.
\begin{table}[h!]
\centering
\begin{tabular}{|l|l|} \hline
 & Reality Condition  \\
 \hline $(0,5)$ &  $(\lambda^i)^* =  B \lambda^j \varepsilon_{j i}$\\
 \hline $(1,4)$ &  $(\lambda^i)^* =  - B \lambda^j \varepsilon_{j i}$\\
 \hline $(2,3)$ &   $(\lambda^i)^* = i B \lambda^j \eta_{i j}$\\ 
 \hline $(3,2)$ &  $(\lambda^i)^* =  - i B \lambda^j \eta_{i j}$\\
 \hline $(4,1)$ &  $(\lambda^i)^* =  B \lambda^j \varepsilon_{j i}$\\
 \hline $(5,0)$ &  $(\lambda^i)^* =  - B \lambda^j \varepsilon_{j i}$
  \\ \hline
\end{tabular}
\caption{Reality Condition in each signature, $B = (C A^{-1})^T$ is
  signature dependent. \label{Values_alpha}}
\end{table}
Since we use the antisymmetric bilinear form $\varepsilon$ on
$\mathbb{C}^2$, the R-symmetry group is
\[
G_R = GL(1,\mathbb{H}') \cap Sp(2,\mathbb{C}) \simeq SU(1,1) \;.
\]

\subsection{Reality properties of spinor bilinears and R-group tensors}

In explicit calculations it is convenient to employ a
formalism where R-symmetry indices $i,j=1,2$ can be raised and
lowered. This is always done using the antisymmetric bilinear form
$\varepsilon$ on $\mathbb{C}^2$, irrespective of the reality
condition. Spinors form R-symmetry doublets, and the rules for 
raising and lowering indices are
\[
\lambda^i \varepsilon_{ij} = \lambda_j \;,\;\;
\lambda^i = \varepsilon^{ij} \lambda_j \;,
\]
where $\varepsilon^{ij}\varepsilon_{jk} = - \delta^i_k$. This rule for
raising and lowering indices conforms with the NW-SE convention. 
Besides spinors, which transform as R-symmetry doublets, the only other
field in a five-dimensional vector multiplet that transforms
non-trivially under R-symmetry is the auxiliary symmetric tensor field
$Y^{ij}$. The rules for raising and lowering indices are
\[
Y^{ij}=\varepsilon^{ik}\varepsilon^{jl} Y_{kl} \;,\;\;\;
Y^{kl} \varepsilon_{ki} \varepsilon_{lj} = Y_{ij} \;.
\]
Spinors and as well the symmetric tensor $Y^{ij}$ are subject to reality
conditions, which involve the indices $i,j$. For spinors we have
\begin{equation}
(\lambda^i)^* = \left\{ \begin{array}{ll}
\alpha B \lambda^j \varepsilon_{ji} & \mbox{for}\; t=0,1,4,5,\\
\alpha B \lambda^j \eta_{ji} & \mbox{for}\; t=2,3,\\
\end{array} \right. \label{Real-Lambda}  \;.
\end{equation}
For symmetric tensors we then have to take the induced reality conditions
\begin{equation}
(Y^{ij})^* = \left\{ \begin{array}{ll}
Y^{kl} \varepsilon_{ki} \varepsilon_{lj} = Y_{ij}\;, &
                                                    \mbox{for}\;\;t=0,1,4,5,\\
Y^{kl} \eta_{ki} \eta_{lj} \;, &
                                                    \mbox{for}\;\;t=2,3\;.\\
\end{array} \right. \label{Real-Y}
\end{equation}
These reality conditions are invariant under the respective R-symmetry
groups $SU(2)$ and $SU(1,1)$,
which act by
\[
Y \rightarrow Y' = U Y U^T \;,
\]
where
\[
U = \left( \begin{array}{cc}
a & b \\
\mp b^* & a^* \\ 
\end{array}\right) \in \left\{ \begin{array}{ll}
SU(2) \; ,&\mbox{for}\;t=0,1,4,5,\\
SU(1,1) ,&\mbox{for}\;t=2,3.\\
\end{array} \right.
\]
Note that $U$ satisfies
\[
U^* = \varepsilon U \varepsilon^T \;,\;\;\;U^*  = \eta U \eta \;,
\]
respectively.\footnote{These relations show that for both groups
the $[2]$ and $[\bar{2}]$ representations are equivalent as real
                                 representations.} 
From this it follows immediately that $Y'$ satisfies the same reality
condition as $Y$.

 Note that for $t=2,3$ $(Y^{ij})^*\not=Y_{ij}$. The explicit
relations between tensor components are:
\[
Y^{11} = Y_{22} \;,\;\;\;
Y^{12} = \pm Y_{12} \;,\;\;\;
Y^{22} = Y_{11} \;,
\]
and
\[
(Y^{11})^* = Y_{11} \;,\;\;\;
(Y^{12})^* = \pm Y_{12} \;,\;\;\;
(Y^{22})^* = Y_{22} \;,
\]
where the plus sign applies for $t=0,1,4,5$ and the minus sign for $t=2,3$.
We also note for later use that $Y^{ij} Y_{ij}$ is real for all signatures.

The final ingredient we need is to determine the reality properties of spinor bilinears
involving real doubled spinors, as these will appear in the
supersymmetry transformations and supersymmetric Lagrangians. 
For the scalar  bilinear, we find
\begin{equation}
\label{lambda-bar-chi}
(\bar{\lambda} \chi)^*= (-1)^t \bar{\lambda}\chi \;.
\end{equation}
This formula only depends on $t$ and does not depend on whether
$\mathbb{S}$ carries a quaternionic or real structure. The reason is
that signs which are sensitive to the type of the reality condition cancel. 
To see this, write out the expression 
\[
(\bar{\lambda} \chi)^*= ((\lambda^i)^T C \chi^j)\varepsilon_{ji})^*
\]
using the reality condition $(\lambda^i)^*= \alpha B \lambda^j
 M_{ji}$, where $M_{ji}=\varepsilon_{ji}$ or $M_{ji} = \eta_{ji}$,
\[
(\bar{\lambda} \chi)^* = - \alpha^2 (\lambda^k)^T B^T C B \chi^i
                                 M_{ki} M_{lj} \varepsilon^{ji} \;.
\]
Using the relations for $A,B,C$ given previously, one finds 
\[
B^T C B = (-1)^{t+1} C \;.
\]
We also need
\[
\varepsilon_{ki} \varepsilon_{lj} \varepsilon^{ji} =
                                                       \varepsilon_{lk}
                                                       \;, \;\;\;
\eta_{ki} \eta_{lj} \varepsilon^{ji} = -  \varepsilon_{lk} \;.
\]
These relations contain a sign which depends on the type of reality condition we
use. However since $\alpha=\pm 1$ for $t=0,1,4,5$, but $\alpha=\pm i$
for $t=2,3$ this sign cancels against $\alpha^2=\pm 1$, and we obtain 
(\ref{lambda-bar-chi}).

Since
\[
(\gamma^{\mu_1 \cdots \mu_l})^* =
(-1)^{tl} B \gamma^{\mu_1 \cdots \mu_l} B^{-1} \;,
\]
this immediately generalizes to 
\begin{equation}
\label{CC-spinor-bilinear}
(\bar{\lambda} \gamma^{\mu_1 \cdots \mu_l} \chi)^* =
(-1)^{t(l+1)} \bar{\lambda} \gamma^{\mu_1 \cdots \mu_l} \chi \;.
\end{equation}
Thus for $t=0$ all spinor bilinears are real,
while for $t=1$ they alternate between imaginary and real. Note that
the vector bilinear, $l=1$, is real for all $t$, as it must, as this is the
bilinear which defines the real supersymmetry algebra.

We conclude by noting that all expressions with completely contracted
auxiliary indices $i,j$ are R-symmetry invariant: Before imposing the
reality condition they are manifestly $Sp(2,\mathbb{C})$-invariant,
and the R-symmetry group is precisely the subgroup of 
$Sp(2,\mathbb{C})$ which commutes with the reality condition.

\subsection{No ${\cal N}=1$ superymmetry algebras in signatures $(2,3)$ and 
$(3,2)$}

For $t=2,3$, the complex spinor module $\mathbb{S}$ carries a 
Spin-invariant real 
structure $J$ defined by the matrix $B$, and we can define Majorana
spinors by imposing $\lambda^* = \alpha B  \lambda$. 
Thus we might be able to define an $N=1$ superalgebra by
restricting the super-admissible bilinear form 
$b=\mbox{Im}(A)$ to $J$-invariant spinors. 
However by explicit computation it is straightforward to verify that 
the Spin$_0$-invariant endomorphism $J:\lambda \mapsto \alpha^*
B^* \lambda^*$ satisfies
\[
b(\gamma^\mu J \lambda, \chi) = - b(\gamma^\mu \lambda, J \chi) 
\]
for all signatures, which implies that the bilinear form is
identically zero when restricted to $J$-invariant spinors. The
corresponding
supersymmetry algebra is trivial, that is, supercharges simply
anti-commute.
Therefore one cannot define an ${\cal N}=1$ supersymmetry algebra in five
dimensions even in those signatures where Majorana spinors exist. 
This conclusion can also be reached by analyzing the tables in 
\cite{Alekseevsky:1997}.

\section{Five-dimensional vector multiplets}

We will now derive the off-shell supersymmetry transformations and 
the general Lagrangians for five-dimensional vector multiplets in
arbitrary signature $(t,s)$. Since we can build on the results of
\cite{Cortes:2003zd} for signature $(1,4)$, we proceed as follows. Initially, we work
on the doubled spinor module  $\mathbb{S}\otimes \mathbb{C}^2$ without
imposing a reality condition. By allowing arbitrary complex
coefficients in the supersymmetry transformations and in the
Lagrangian known from \cite{Cortes:2003zd}, we obtain all conditions on these
parameters which are independent of signature. The resulting
transformations and Lagrangian can be viewed as a common
complexification of all the five-dimensional theories. In a second
step we impose the appropriate reality conditions for each
signature,  and obtain further signature dependent conditions on the
parameters. We will show that consistent solutions exist for all
signatures, and are unique up to conventional choices. The
transformations and Lagrangians vary from signature to signature only 
by sign flips or insertion of factors of $\pm i$. Moreover, theories 
in signatures $(t,s)$ and $(s,t)$ are related by changing the
convention for the space-time metric from `mostly plus' to `mostly
minus', plus taking into account changes in the reality properties of
spinor bilinears. Thus we end up with three physically distinct cases:
Euclidean theories with signatures $(0,5)$ and $(5,0)$, 
Lorentzian theories with signatures $(1,4)$ and $(4,1)$ and
exotic theories with two time-like directions, with signatures
$(2,3)$ and $(3,2)$.

\subsection{Supersymmetry variations}

We have seen that when using doubled spinors 
the supersymmetry algebra to takes the form
\[
\{ Q_{\alpha i},Q_{\alpha j} \} = -\frac{1}{2} (\gamma^\mu
C^{-1})_{\alpha \beta} P_\mu \varepsilon_{ij}  \;.
\]
To show that this algebra is represented on a collection of
space-time fields, we must verify that 
\begin{equation}
\label{QQonFields}
[\bar{\epsilon}_{(1)} Q, \bar{\epsilon}_{(2)} Q ]  \Phi(x) =
-\frac{1}{2} (\bar{\epsilon}_{(1)} \gamma^\mu C^{-1}
\epsilon_{(2)}) \partial_\mu \Phi(x)
\end{equation}
for all fields $\Phi(x)$ belonging to the given supermultiplet.
The supersymmetry parameters $\epsilon_{(i)}$ and the
fermionic fields $\lambda^i$ are anti-commuting spinor
fields.\footnote{Note that so far we have been using commuting
  spinors. Going from commuting to anti-commuting spinors is a
  `natural' (functorial) operation and introduces 
signs in any operation which involves changing the order of
spinors. }
In implementing the supersymmetry algebra, we have chosen that
the translation operator $P_\mu$ acts as $\partial_\mu$ on fields, which is the same 
convention as in \cite{Freedman:2012zz}.\footnote{In the mathematics 
literature translations are taken to act by $-\partial_\mu$ on functions, since
this is the infinitesimal action of the one-parameter group
$\varphi_t\;:{\bf x} \mapsto {\bf x} + t {\bf e}_\mu$ of
diffeomorphisms on functions. }

The content of a five-dimensional off-shell vector multiplets 
is known from signature $(1,4)$:
\begin{equation} (A^{\mu}, \lambda^i,  \sigma, Y^{i j}) \;,\qquad \mu
  = 1,2,3,4,5 \;,  \qquad i = 1, 2 \;.
\end{equation}
Apart from the eponymous vector field, $A^{\mu}$, a vector multiplet
contains  a pair of spinors, $\lambda^i$, a scalar field, $\sigma$,
and a triplet of auxiliary field combined into  a symmetric tensor, $Y^{ij}$. 
At this point no reality conditions are imposed on $\lambda^i$ and
$Y^{ij}$, and therefore $A^\mu$ and $\sigma$ should be considered as
complex fields in
order to balance bosonic and fermionic degrees of freedom. 

Since we anticipate that the numerical coefficients of terms in the
supersymmetry variations will depend on the reality conditions we will
impose later,  we take the off-shell supersymmetry transformations derived in
\cite{Cortes:2003zd} for signature $(1,4)$, but replace the numerical
pre-factors of all terms by complex coefficients:
\begin{eqnarray} \delta A^{\mu} &=& \alpha \bar{\epsilon} \gamma^{\mu} \lambda \;,\qquad  \delta \sigma = a \bar{\epsilon} \lambda \;,\qquad \delta Y^{i j} =  u \bar{\epsilon}^{(i}\cancel{\partial} \lambda^{j)} \;, \label{transys} \\
\delta \lambda^{i} &=& \beta \gamma \cdot F \epsilon^i +b \cancel{\partial} \sigma \epsilon^i +  y Y^{i j} \epsilon_j \;.\nonumber
\end{eqnarray}
By imposing that the transformations satisfy  (\ref{QQonFields}), and
therefore realize the complex supersymmetry algebra,
we obtain the following independent relations between
the coefficients:
\begin{align} &- \frac{1}{2} = - 2ab = 4 \alpha \beta =  - u y\;. \label{xirel}
\end{align}
We remark that the calculation leading to these relations
is identical to the one done previously in \cite{Cortes:2003zd}, except that we have replaced numerical coefficients by
parameters.

The relations (\ref{xirel}) can be solved for  $b,\beta, y$ in terms of
$a,\alpha,u$:
\[
b=\frac{1}{4a} \;,\;\;\;\beta=-\frac{1}{8\alpha} \;,\;\;\;
y = \frac{1}{2u} \;.
\]


We now impose reality conditions, proceeding signature by signature. For the scalar
$\sigma$ and the vector $A^\mu$ this simply means that they are now
real-valued. The reality conditions imposed on the spinors
$\lambda^i$, $\epsilon^i$ were given in (\ref{Real-Lambda}) while
those imposed on the auxiliary tensor $Y^{ij}$ were given in
(\ref{Real-Y}). With regard to the reality properties of spinor bilinears we
need to take into account that in a field theory realisation
spinors are anticommuting quantities. This leads to extra minus signs
when applying operations which change the order of spinors, such as
transposition or Hermitian conjugation. Complex conjugation can either
be defined to change the order of anticommuting fields or not, and both
conventions can be
mapped to each other by changing  the phase factor in the reality
condition \cite{VanProeyen:1999ni,Bergshoeff:2001hc}.\footnote{This 
phase factor was denoted 
$\alpha$ previously, but is of course distinct from the parameter
$\alpha$ in the supersymmetry transformations. Since by now the
reality conditions have been fixed according to Table
\ref{Values_alpha} this should not
cause any confusion.}
For us it is convenient to keep the reality conditions we fixed in the
previous section, and therefore we define the complex conjugation of
anti-commuting spinors such that it does 
not change the order.
Then we can use (\ref{CC-spinor-bilinear}) to work
out the reality properties of the parameters $\alpha,a$ by imposing
that $\delta \sigma$ and $\delta A^\mu$ are real.
Similary, by imposing reality on the variations $\delta Y^{ij}$ and
$\delta \lambda^i$, we obtain the reality conditions for
$u, \beta, b,y$. The resulting conditions are summarized in Table
\ref{Susy-parameters-real-1}.

\begin{table}[h!]
\centering
\begin{tabular}{|l|l|l|} \hline
{Parameter} & {Real} & {Imaginary} \\ \hline
$\alpha$  & $t=0, \ldots 5$ &  {never} \\ \hline
$a$ & $t=0,2,4$& $t=1,3,5$ \\ \hline
$\beta$ & $t=0, \ldots, 5$ & {never} \\ \hline
$b$ & $t=0,2,4$ & $t=1,3,5$ \\ \hline
$u$ & $t=0,1,4,5$ & $t=2,3$ \\ \hline
$y$ & $t=0,1,4,5$ & $t=2,3$ \\ \hline
\end{tabular}
\caption{Reality conditions for the coefficients of the supersymmetry
transformations. \label{Susy-parameters-real-1}} 
\end{table}

These conditions are consistent with (\ref{xirel}), because the
products $ab, \alpha \beta, uy$ are real for all signatures. 
For signature $(1,4)$ the resulting reality properties of
symplectic Majorana spinor bilinears and supersymmetry variations 
agree with \cite{Bergshoeff:2001hc} and \cite{Cortes:2003zd}. For concreteness, we now make 
explicit choices for $a,\alpha,u$, which then determine $b,\beta,y$
through (\ref{xirel}). For $(1,4)$ we make a choice which 
reproduces \cite{Cortes:2003zd}. For other signatures we then introduce 
factors $-1, \pm i$, where needed. 
We summarize the resulting values for the parameters in Table
\ref{Susy-parameters-real}.

\begin{table}[h!]
\centering
\begin{tabular}{|c|c|c|c|c|c|c|} \hline
 & $\alpha$ & $a$ & $u$ & $\beta$ & $b$ & $y$ \\ \hline
$(0,5)$ & $\frac{1}{2}$ & $\frac{1}{2}$ & $-\frac{1}{2}$ &
                                                           $-\frac{1}{4}$
                                  & $\frac{1}{2}$ & $-1$ \\ \hline
$(1,4)$ & $\frac{1}{2}$ & $\frac{i}{2}$ & $-\frac{1}{2}$ &
                                                           $-\frac{1}{4}$
                                  & $-\frac{i}{2}$ & $-1$ \\ \hline
$(2,3)$ & $\frac{1}{2}$ & $\frac{1}{2}$ & $-\frac{i}{2}$ &
                                                           $-\frac{1}{4}$
                                  & $\frac{1}{2}$ & $i$ \\ \hline
$(3,2)$ & $\frac{1}{2}$ & $\frac{i}{2}$ & $-\frac{i}{2}$ &
                                                           $-\frac{1}{4}$
                                  & $-\frac{i}{2}$ & $i$ \\ \hline
$(4,1)$ & $\frac{1}{2}$ & $\frac{1}{2}$ & $-\frac{1}{2}$ &
                                                           $-\frac{1}{4}$
                                  & $\frac{1}{2}$ & $-1$ \\ \hline
$(5,0)$ & $\frac{1}{2}$ & $\frac{i}{2}$ & $-\frac{1}{2}$ &
                                                           $-\frac{1}{4}$
                                  & $-\frac{i}{2}$ & $-1$ \\ \hline
\end{tabular}
\caption{Explicit choices for the parameters in the supersymmetry
  transformations for all signatures. \label{Susy-parameters-real}} 
\end{table}

\subsection{Supersymmetric Lagrangians}

The representations we have found above are off-shell and therefore
independent of a choice of field equations or Lagrangian. To specify 
Lagrangians we proceed in the same way as with the supersymmetry
transformations. 
In \cite{Cortes:2003zd} the general (two derivative) rigid off-shell Lagrangian for vector
multiplets in signature $(1,4)$ was determined. To adapt this result to
other signatures, we introduce parameters $s_F, s_\sigma, s_\lambda,
s_Y, \theta_1, \theta_2, \theta_3 $, which are fixed
later by imposing reality conditions. 

The general form of the Lagrangian is
\begin{align} L = &\bigg( \frac{s_F}{4} F^I _{\mu \nu} F^{J \mu \nu} + \frac{s_{\sigma}}{2} \partial_{\mu} \sigma^I \partial^{\mu} \sigma^J + \frac{s_{\lambda}}{2} \bar{\lambda}^I \cancel{\partial} \lambda^J + s_Y Y^I_{i j} Y^{J i j } \bigg) \mathcal{F}_{I J} (\sigma) \\
&+ \bigg( \theta_1 \varepsilon^{\mu \nu \rho \sigma \tau} A^{I} _{\mu} F^{J} _{\nu \rho} F^{K} _{\sigma \tau} + \theta_2 \bar{\lambda}^I \gamma^{\mu \nu} F^J _{\mu \nu}\lambda^K  + \theta_3 \bar{\lambda}^{I i} \lambda^{J j} Y^K_{i j} \bigg) {\cal{F}} _{I J K} (\sigma) \;.\nonumber
\end{align}
Vector multiplets are labeled by  $I,J = 1, ..., N$. Initially
we do not impose reality conditions, so that all quantities are
complex. Note that this Lagrangian is complex-valued, and holomorphic,
since it does not involve complex conjugation. As in \cite{R-symmetry}
it does not seem to have a direct physical interpretation\footnote{Except possibly 
in terms of a complexified configuration space which contains the complex 
saddle points of a Euclidean functional integral, see for example
\cite{Mohaupt:2009iq,Mohaupt:2010du}.} but can be thought of as a 
`holomorphic master Lagrangian,' encoding all the possible real forms.
Imposing invariance
under the supersymmetry transformations (\ref{transys}) involves the
same computational steps as verifying the supersymmetry invariance 
of the $(1,4)$ theory in \cite{Cortes:2003zd}. Therefore the essential
structural  properties carry over to all signatures. In particular, 
the scalar field dependent couplings ${\cal F}_{IJ}$ and ${\cal
  F}_{IJK}$ can be expressed as derivatives of a Hesse potential 
${\cal F}(\sigma)$:\footnote{As long as we do not impose reality
  conditions this is a holomorphic Hesse potential for a
  complex-Riemannian metric.} 
\begin{align}
	\mathcal{F}_{I J} (\sigma)= \frac{\partial}{\partial \sigma^I}
  \frac{\partial}{\partial \sigma^J}\mathcal{F}(\sigma) \;,\qquad
  \mathcal{F}_{I J K}(\sigma) = \frac{\partial}{\partial \sigma^I}
  \frac{\partial}{\partial \sigma^J}\frac{\partial}{\partial
  \sigma^K}\mathcal{F}(\sigma) \;.
\end{align} 
Supersymmetry requires that a Chern-Simons term is included, unless
the theory is non-interacting and thus the Hesse potential quadratic.
Gauge invariance of the Chern-Simons term, up to boundary terms, 
implies that ${\cal F}(\sigma)$ must be a polynomial of degree at most three.
As noted in \cite{Cortes:2003zd} the spinor term can be written using a
partial derivative, rather than using the covariant derivative with respect
to the Levi-Civita connection of the scalar manifold, because the 
difference is identically zero due to the property 
$\bar{\lambda}^{i (I|} \lambda^{j| J)} \varepsilon_{j i} = 0$ of
spinor bilinears, which holds irrespective of reality conditions.


Invariance of the Lagrangian under the transformations \eqref{transys}
imposes the following conditions on the coefficients:
\begin{align} &s_F \alpha = - 2 s_{\lambda} \beta\;, \quad
  &&s_{\sigma} a = 
- s_{\lambda} b\;, \quad &&&2 s_Y u = -  s_{\lambda} y \;,\nonumber \\
& 3 \theta_1 \alpha = \pm 2 i^t \theta_2 \beta\;, \quad  &&4  \theta_2
                                                          \alpha = -
                                                          \theta_3 u\;,
                                                          \quad
                                                                                               &&&\theta_2
                                                                                                   y
                                                                                                   =
                                                                                                   \theta_3
                                                                                                   \beta
                                                                                                   \;, \label{Invariance-Lagrangian}\\
&\alpha s_F = 8 b \theta_2 \;,\quad &&a s_Y = y \theta_3 \;,\quad &&&a s_{\lambda} = 8 \alpha \theta_2 \;.\nonumber
\end{align}
The sign appearing in the relation determining $\theta_1$ will be
explained below.

The relations (\ref{Invariance-Lagrangian}) are consistent but not
independent. Using the relations (\ref{xirel}) we can rearrange
the condition imposed on the coefficients $s_i$ as follows:
\begin{align} - \frac{1}{2} = - 2 \frac{s_F}{s_{\lambda}} \alpha^2= 2 \frac{s_{\sigma}}{s_{\lambda}} a^2 = 2 \frac{s_Y}{s_{\lambda}} u^2 \label{relates}
\end{align}
These relations determine the relative signs of the quadratic terms in
the Lagrangian. The coefficients $\theta_2, \theta_3$ are then determined
in terms of any of the $s_i$, say $s_\lambda$ for concreteness,  as:
\[
\theta_2 = \frac{a}{8\alpha} s_\lambda \;,\;\;\;
\theta_3 = - \frac{9}{16 u} s_\lambda  \;.
\]
Finally, $\theta_1$ is determined by
\[
\theta_1 = \pm \frac{2}{3} i^t \theta_2 \frac{\beta}{\alpha} = 
\pm \frac{2}{3} i^t \frac{a\beta}{8 \alpha^2} s_\lambda
=\pm \frac{1}{24} \;,
\]
with a sign that can be adjusted by making conventional choices.
In odd dimensions the volume element 
$\omega = \gamma_1 \cdots \gamma_{t+s}$ of the Clifford algebra 
commutes with the generators $\gamma_\mu$ and is therefore 
proportional to the unit $\mathbbm{1}$ on irreducible representations.
Since in five dimensions $\omega^2 = (-1)^t \mathbbm{1}$, this implies
that $\omega = \pm \dblone$ for even $t$ and $\omega = \pm i \dblone$
for odd $t$. 
The choice of a sign, together with the definition 
of the completely antisymmetric tensor $\varepsilon_{\mu \nu \rho
  \sigma \tau}$ determines the relative factor in the relation 
$\gamma_{\mu \nu \rho \sigma \tau} \propto \varepsilon_{\mu \nu \rho
  \sigma \tau}  \mathbbm{1}$
for completely antisymmetrized products of five generators. 
The choices 
we make are $\omega = \mathbbm{1}$ for even $t$, 
$\omega = -i \mathbbm{1}$ for odd $t$ and $\varepsilon_{01234} = +1$
for all signatures. Then
\[
\gamma_{\mu \nu \rho \sigma \tau} = \left\{ 
\begin{array}{cc}
\varepsilon_{\mu \nu \rho \sigma \tau} \dblone\;, & t \;\;\mbox{even}\;, \\
- i \varepsilon_{\mu \nu \rho \sigma \tau} \dblone \;,& t
                                                        \;\;\mbox{odd}
  \;.\\
\end{array} \right.
\]
These choices are relevant when relating the $\gamma$ and 4--$\gamma$
terms arising from the supersymmetry 
variations of the Chern Simons term and of the
term $\bar{\lambda}^I\gamma^{\mu \nu} F_{\mu \nu}^J
\lambda^K$. With our conventions the Lagrangian is supersymmetric 
for all signatures if $\theta_1 = + \frac{1}{24}$.


The values of $\alpha^2, a^2, u^2$ depend on the signature, leading
to sign flips of the respective terms between signatures. For 
reference we will now list the Lagrangians and supersymmetry 
transformations for all signatures. 

\subsection{Overview and discussion of supersymmetry variations and
  Lagrangians by signature}

\subsubsection{Signature $(1,4)$}

In signature $(1,4)$, the
relations (\ref{relates}) imply:
$s_{\sigma} = s_{\lambda} =  s_F = - s_Y =\pm 1$.
Thus all physical fields have the same sign in front of their kinetic
terms. While the overall sign of the Lagrangian cannot be fixed by
imposing invariance under supersymmetry, there is a standard choice
in Minkowski signature, namely the choice which makes the 
kinetic terms of all physical fields positive definite.
Therefore we choose 
\begin{align}
& s_{\sigma} = s_{\lambda} =  s_F = - s_Y =- 1\;.
\end{align}

The remaining coefficients can then be determined from 
(\ref{Invariance-Lagrangian}). We find
\[
\theta_2 = \frac{i}{8} s_F = -\frac{i}{8} \;,\;\;\;
\theta_3 = -\frac{i}{2} s_Y = -\frac{i}{2} \;.
\]
Finally, as explained above, we have made conventional choices such
that $\theta_1= \frac{1}{24}$ for all signatures.

Having determined all coefficients, we now summarize the resulting 
$(1,4)$ signature
Lagrangian and supersymmetry transformations:
\begin{align}
L_{(1,4)} = &\bigg( - \frac{1}{4} F^I_{\mu \nu} F^{J \mu \nu} - \frac{1}{2} \partial_{\mu} \sigma^I \partial^{\mu} \sigma^J - \frac{1}{2} \bar{\lambda}^I \cancel{\partial} \lambda^J + Y^I_{i j} Y^{i j J} \bigg) \mathcal{F}_{I J} \\
&+ \bigg( \frac{1}{24} \varepsilon^{\mu \nu \rho \sigma \tau} A^{I}
  _{\mu} F^{J} _{\nu \rho} F^{K} _{\sigma \tau} - \frac{i}{8}
  \bar{\lambda}^I \gamma^{\mu \nu} F^J _{\mu \nu} \lambda^K  -
  \frac{i}{2} \bar{\lambda}^{I i} \lambda^{J j} Y^K_{i j} \bigg)
  {\cal{F}} _{I J K} \;.\nonumber \\ \delta A^I_{\mu} &=   \frac{1}{2}
                                                        \bar{\epsilon}
                                                        \gamma_{\mu}
                                                        \lambda^I \;,
                                                        \qquad  \delta
                                                        \sigma^I =
                                                        \frac{i}{2}
                                                        \bar{\epsilon}
                                                        \lambda^I \;,
                                                        \qquad \delta
                                                        Y^{i j I} = -
                                                        \frac{1}{2}
                                                        \bar{\epsilon}^{(i}\cancel{\partial}
                                                        \lambda^{j) I}
                                                        \;, \\
\delta \lambda^{i I} &=  - \frac{1}{4} \gamma^{\mu \nu} F^I _{\mu \nu}
                       \epsilon^i - \frac{i}{2} \cancel{\partial}
                       \sigma^I \epsilon^i  - Y^{i j I} \epsilon_j
                       \;. \nonumber
\end{align}
These results agree with \cite{Cortes:2003zd}. Note that in \cite{Cortes:2003zd} the 
Chern-Simons terms appears with the opposite sign, because there the 
$\varepsilon$-tensor was defined with a relative minus sign, compared
to our definition above.

Since the scalar fields are now real, we recover the well known
affine special real geometry of five-dimensional vector multiplets,
where the scalar metric is a Hessian metric with a cubic polynomial as
Hesse potential. This cubic potential needs to be chosen such that
the resulting metric is positive definite, a condition that needs to 
investigated case by case.

\subsubsection{Signature $(4,1)$}

Next consider signature $(4,1)$, which we interpret as Minkowski
signature with a mostly minus convention for the metric. Now
(\ref{relates}) implies
\begin{align}
& -s_{\sigma} = s_{\lambda} =  s_F = - s_Y =\pm 1\;.
\end{align}
This time the scalar and vector kinetic have different signs, but this
only reflects that we now use a different convention for the metric.
The standard choice where all physical fields have a positive definite
kinetic term is
\begin{align}
& -s_{\sigma} = s_{\lambda} =  s_F = - s_Y =- 1\;.
\end{align}
From this we determine\footnote{In the following it is always
  understood that we chosen the convention which makes $\theta_1$
  positive, as explained above.}
\[
\theta_1 = \frac{1}{24}  \;,\;\;\;\theta_2 = -\frac{1}{8} \;,\;\;\;
\theta_3 =  - \frac{1}{2} \;.
\]
The resulting Langrangian and supertransformations are:
\begin{align}
L_{(4,1)} = &\bigg( - \frac{1}{4} F^I_{\mu \nu} F^{J \mu \nu} + \frac{1}{2} \partial_{\mu} \sigma^I \partial^{\mu} \sigma^J - \frac{1}{2} \bar{\lambda}^I \cancel{\partial} \lambda^J + Y^I_{i j} Y^{i j J} \bigg) \mathcal{F}_{I J} \\
&+ \bigg(\frac{1}{24} \varepsilon^{\mu \nu \rho \sigma \tau} A^{I} _{\mu} F^{J} _{\nu \rho} F^{K} _{\sigma \tau} - \frac{1}{8} \bar{\lambda}^I \gamma^{\mu \nu} F^J _{\mu \nu} \lambda^K  - \frac{1}{2} \bar{\lambda}^{I i} \lambda^{J j} Y^K_{i j} \bigg) {\cal{F}} _{I J K} \;.\nonumber \\ \delta A^I_{\mu} &=   \frac{1}{2} \bar{\epsilon} \gamma_{\mu} \lambda^I \;,\qquad  \delta \sigma^I = \frac{1}{2} \bar{\epsilon} \lambda^I \;,\qquad \delta Y^{i j I} = - \frac{1}{2} \bar{\epsilon}^{(i}\cancel{\partial} \lambda^{j) I} \;,\\
\delta \lambda^{i I} &=  - \frac{1}{4} \gamma^{\mu \nu} F^I _{\mu \nu} \epsilon^i + \frac{1}{2} \cancel{\partial} \sigma^I \epsilon^i  - Y^{i j I} \epsilon_j \;.\nonumber
\end{align}
When comparing the signatures $(1,4)$ and $(4,1)$ we observe that they
differ by factors $\pm 1$ and $\pm i$, which can be interpreted as
resulting from changing from a `mostly plus' to a `mosly minus'
convention for the metric. For some of the fermionic terms this
involves a factor of $i$ due to the signature dependence of the reality
properties of spinor bilinears. Interpreting the single distinguished direction
as time, both theories are equivalent.

\subsubsection{Signature $(0,5)$}

Now we turn to Euclidean signatures. For signature $(0,5)$, 
(\ref{relates}) implies
\begin{align}
& -s_{\sigma} = s_{\lambda} =  s_F = - s_Y =\pm 1\;.
\end{align}
We observe that the Euclidean action has a relative sign between
the scalar and vector kinetic term, and therefore cannot be 
positive definite. In \cite{Sabra:2016abd} the existence of
five-dimensional  Euclidean vector multiplet 
actions with a relative sign between scalar and vector term
was deduced using dimensional 
reduction of the bosonic Lagrangians of higher-dimensional
supergravity theories, and of their Killing spinor equations. We have now
derived this result by imposing supersymmetry directly in five
dimensions. Our derivation shows in particular that there is no
option, because this relative sign is required by supersymmetry. 
We remark that this is a non-trivial insight, since, for instance,
it was argued in \cite{Sabra:2016abd} 
that an analogous relative sign for four-dimensional Euclidean
vector multiplets can be changed by a field redefinition. We 
will present a full off-shell analysis of the four-dimensional
${\cal N}=2$ vector multiplet theories  in \cite{Euc4d}.

Since the supersymmetric Euclidean action is not definite, there is 
no preferred choice of a sign. We make the conventional choice
$s_\lambda=-1$ which results in
\begin{align}
L_{(0,5)} = &\bigg( - \frac{1}{4} F^I_{\mu \nu} F^{J \mu \nu} + \frac{1}{2} \partial_{\mu} \sigma^I \partial^{\mu} \sigma^J - \frac{1}{2} \bar{\lambda}^I \cancel{\partial} \lambda^J + Y^I_{i j} Y^{i j J} \bigg) \mathcal{F}_{I J} \\
&+ \bigg(\frac{1}{24} \varepsilon^{\mu \nu \rho \sigma \tau} A^{I}
  _{\mu} F^{J} _{\nu \rho} F^{K} _{\sigma \tau} - \frac{1}{8}
  \bar{\lambda}^I \gamma^{\mu \nu} F^J _{\mu \nu} \lambda^K  -
  \frac{1}{2} \bar{\lambda}^{I i} \lambda^{J j} Y^K_{i j} \bigg)
  {\cal{F}} _{I J K} \;.\nonumber \\ \delta A^I_{\mu} &=   \frac{1}{2}
                                                        \bar{\epsilon}
                                                        \gamma_{\mu}
                                                        \lambda^I
                                                        \;,\qquad
                                                        \delta
                                                        \sigma^I =
                                                        \frac{1}{2}
                                                        \bar{\epsilon}
                                                        \lambda^I \;,
                                                        \qquad \;,
                                                        \delta Y^{i j
                                                        I} = -
                                                        \frac{1}{2}
                                                        \bar{\epsilon}^{(i}\cancel{\partial}
                                                        \lambda^{j) I}
  \;, \\
\delta \lambda^{i I} &=  - \frac{1}{4} \gamma^{\mu \nu} F^I _{\mu \nu}
                       \epsilon^i + \frac{1}{2} \cancel{\partial}
                       \sigma^I \epsilon^i  - Y^{i j I} \epsilon_j \;. \nonumber
\end{align}

\subsubsection{Signature $(5,0)$}

For signature $(5,0)$ we find instead
\begin{align}
& s_{\sigma} = s_{\lambda} =  s_F = - s_Y =\pm 1\;.
\end{align}
Taking into account that the space-time metric is negative definite,
the Euclidean action is again indefinite. As for signature $(0,5)$ we  make a conventional
choice of the overall sign:
\begin{align}
L = &\bigg( - \frac{1}{4} F^I_{\mu \nu} F^{J \mu \nu} - \frac{1}{2} \partial_{\mu} \sigma^I \partial^{\mu} \sigma^J - \frac{1}{2} \bar{\lambda}^I \cancel{\partial} \lambda^J + Y^I_{i j} Y^{i j J} \bigg) \mathcal{F}_{I J} \\
&+ \bigg( \frac{1}{24} \varepsilon^{\mu \nu \rho \sigma \tau} A^{I}
  _{\mu} F^{J} _{\nu \rho} F^{K} _{\sigma \tau} - \frac{i}{8}
  \bar{\lambda}^I \gamma^{\mu \nu} F^J _{\mu \nu} \lambda^K  -
  \frac{i}{2} \bar{\lambda}^{I i} \lambda^{J j} Y^K_{i j} \bigg)
  {\cal{F}} _{I J K} \;.\nonumber \\ \delta A^I_{\mu} &=   \frac{1}{2}
                                                        \bar{\epsilon}
                                                        \gamma_{\mu}
                                                        \lambda^I
                                                        \;,\qquad
                                                        \delta
                                                        \sigma^I =
                                                        \frac{i}{2}
                                                        \bar{\epsilon}
                                                        \lambda^I 
\;, \qquad \delta Y^{i j I} = - \frac{1}{2}
                                                        \bar{\epsilon}^{(i}\cancel{\partial}
                                                        \lambda^{j) I}
                                                        \;,  \\
\delta \lambda^{i I} &=  - \frac{1}{4} \gamma^{\mu \nu} F^I _{\mu \nu} \epsilon^i - \frac{i}{2} \cancel{\partial} \sigma^I \epsilon^i  - Y^{i j I} \epsilon_j \;.\nonumber
\end{align}

\subsubsection{Signature $(2,3)$}

We now turn to the exotic signatures with two time-like directions. 
In signature $(2,3)$, (\ref{relates}) implies
\begin{align}
& -s_{\sigma} = s_{\lambda} =  s_F =  s_Y =\pm 1\;.
\end{align}
With a conventional choice of overall sign, we obtain:
\begin{align}
L_{(2,3)} = &\bigg( - \frac{1}{4} F^I_{\mu \nu} F^{J \mu \nu} + \frac{1}{2} \partial_{\mu} \sigma^I \partial^{\mu} \sigma^J - \frac{1}{2} \bar{\lambda}^I \cancel{\partial} \lambda^J - Y^I_{i j} Y^{i j J} \bigg) \mathcal{F}_{I J} \\
&+ \bigg( \frac{1}{24} \varepsilon^{\mu \nu \rho \sigma \tau} A^{I}
  _{\mu} F^{J} _{\nu \rho} F^{K} _{\sigma \tau} - \frac{1}{8}
  \bar{\lambda}^I \gamma^{\mu \nu} F^J _{\mu \nu} \lambda^K  +
  \frac{i}{2} \bar{\lambda}^{I i} \lambda^{J j} Y^K_{i j} \bigg)
  {\cal{F}} _{I J K} \;, \nonumber \\ \delta A^I_{\mu} &=
                                                         \frac{1}{2}
                                                         \bar{\epsilon}
                                                         \gamma_{\mu}
                                                         \lambda^I\;,
                                                         \qquad
                                                         \delta
                                                         \sigma^I =
                                                         \frac{1}{2}
                                                         \bar{\epsilon}
                                                         \lambda^I \;,
                                                         \qquad \delta
                                                         Y^{i j I} = -
                                                         \frac{i}{2}
                                                         \bar{\epsilon}^{(i}\cancel{\partial}
                                                         \lambda^{j)
                                                         I} \;, \\
\delta \lambda^{i I} &=  - \frac{1}{4} \gamma^{\mu \nu} F^I _{\mu \nu} \epsilon^i + \frac{1}{2} \cancel{\partial} \sigma^I \epsilon^i  + i Y^{i j I} \epsilon_j \;.\nonumber
\end{align}

\subsubsection{Signature $(3,2)$}

Finally, in signature $(3,2)$ we find
\begin{align}
& s_{\sigma} = s_{\lambda} =  s_F =  s_Y =\pm 1\;,
\end{align}
and with a conventional choice of overall sign
\begin{align}
L_{(3,2)} = &\bigg( - \frac{1}{4} F^I_{\mu \nu} F^{J \mu \nu} - \frac{1}{2} \partial_{\mu} \sigma^I \partial^{\mu} \sigma^J - \frac{1}{2} \bar{\lambda}^I \cancel{\partial} \lambda^J - Y^I_{i j} Y^{i j J} \bigg) \mathcal{F}_{I J} \\
&+ \bigg( \frac{1}{24} \varepsilon^{\mu \nu \rho \sigma \tau} A^{I}
  _{\mu} F^{J} _{\nu \rho} F^{K} _{\sigma \tau} - \frac{i}{8}
  \bar{\lambda}^I \gamma^{\mu \nu} F^J _{\mu \nu} \lambda^K  -
  \frac{1}{2} \bar{\lambda}^{I i} \lambda^{J j} Y^K_{i j} \bigg)
  {\cal{F}} _{I J K} \;, \nonumber \\ \delta A^I_{\mu} &=
                                                         \frac{1}{2}
                                                         \bar{\epsilon}
                                                         \gamma_{\mu}
                                                         \lambda^I \;,
                                                         \qquad
                                                         \delta
                                                         \sigma^I =
                                                         \frac{i}{2}
                                                         \bar{\epsilon}
                                                         \lambda^I
                                                         \;, \qquad
                                                         \delta Y^{i j
                                                         I} = -
                                                         \frac{i}{2}
                                                         \bar{\epsilon}^{(i}\cancel{\partial}
                                                         \lambda^{j)
                                                         I} \;, \\
\delta \lambda^{i I} &=  - \frac{1}{4} \gamma^{\mu \nu} F^I _{\mu \nu} \epsilon^i - \frac{i}{2} \cancel{\partial} \sigma^I \epsilon^i  + i Y^{i j I} \epsilon_j\;. \nonumber
\end{align}
While the interpretation of relative and overall signs is not obvious
for a theory with multiple time-like directions, we observe that
all relative sign flips between signatures $(2,3)$ and $(3,2)$ can 
be interpreted as going from a mostly plus to a mostly minus convention
for the space-time metric, together with factors $i$ to account for
the reality properties of spinor bilinears. 

We finally note that by comparing our results signature by signature
to \cite{Sabra:2017xvx} it is straightforward to check that they agree 
where comparable, that is for the bosonic terms of the Lagrangians and 
the supersymmetry variations of the fermions.

\section{Open problems and Outlook}

In this paper we have taken the first step to developing a formalism
which allows us to construct supersymmetric theories simultanously 
for all space-time signatures. We have restricted ourselves to
supersymmetry algebras based on complex irreducible spinor
representations, where we determined the possible R-symmetry groups. 
In a companion paper we will extend this analysis to the general 
case \cite{R-symmetry}. As an application we constructed off-shell
vector multiplets and the associated supersymmetric Lagrangians for
the unique minimal five-dimensional supersymmetry algebra in arbitrary
signature. The five dimensional case is particularly straightforward
to fully analyze, because (i) Dirac spinors are complex
irreducible, (ii) Majorana conditions do not lead
to a non-trivial smaller supersymmetry algebra, and (iii) the space of
superbrackets is one-dimensional. The natural next application, rigid
four-dimensional ${\cal N}=2$ vector multiplets, to be presented in 
\cite{Euc4d}, is much richer. Four-dimensional Dirac spinors are complex
reducible, and in some signatures a Majorana condition defines a
smaller supersymmetry algebra. Moreover, from 
\cite{Alekseevsky:1997} one can read off that the space of ${\cal
  N}=2$ superbrackets is four-dimensional, and it is not clear 
a priori whether all these superbrackets define isomorphic Lie
superalgebras. The scalar geometry of 
four-dimensional vector multiplets depends on the signature:
it is special K\"ahler 
in Lorentz signature, but special para-K\"ahler in Euclidean signature
\cite{Cortes:2003zd}. This makes the issues concerning relative signs
between kinetic terms in the Lagrangian much more interesting. 
In \cite{Sabra:2016abd} it was observed that 
by dimensional reduction one can arrive at two different formulations
of the bosonic sector of four-dimensional Euclidean vector multiplets, 
which differ by a relative sign flip between the scalar and the vector
term. While it was shown in \cite{Sabra:2016abd} that both
formulations of the bosonic sector are related by a field
redefinition, it remains to be  seen
whether and how this extends to the fermionic terms and supersymmetry 
transformations. Also according to \cite{Sabra:2017xvx} the reduction 
of exotic five-dimensional theories to Lorentzian signature gives rise
to non-standard signs for some kinetic terms, similar to type II$^*$
theories in ten dimensions. In \cite{Euc4d} we will provide a complete
analysis, based on a further development of the formalism presented in
this paper. Other directions which we will explore is the application of our
formalism to other multiplets, notably hypermultiplets and the Weyl
multiplet. The later will allow us to construct off-shell realizations
of ${\cal N}=2$ supergravity in arbitrary signature.

\subsection*{Acknowledgements}

T.M. thanks Vicente Cort\'es for many useful discussions and the 
Department of Mathematics and Centre for Mathematical Physics
of the University of Hamburg for support and hospitality during
various stages of this work. The work of T.M. was partly supported by the
STFC consolidated grant ST/G00062X/1. The work of L.G. was supported
by STFC studentship ST/1643452.

\begin{appendix}
\section{Spinor conventions \label{Spinor_Con}}

We use the same conventions for spinor indices as in \cite{Cortes:2003zd}. 
Dirac spinors $\psi \in \mathbb{S}$ have lower indices, $\psi=(\psi_\alpha)$.
$\gamma$ matrices are endomorphisms on the spinor module, 
$\gamma^\mu= (\gamma^{\mu\;\;\beta}_{\;\;\alpha})$. The matrices $A$
and $C$ represent a sesquilinear form and a
bilinear form on $\mathbb{S}$: $A=(A^{\alpha\beta})$, and
$C=(C^{\alpha \beta})$. The inverse matrices are denoted
$A^{-1}=(A_{\alpha \beta})$, and $C^{-1}=(C_{\alpha \beta})$. 
Note that the definition $A = \gamma_1 \cdots \gamma_t$ is 
an equation between matrices, not between maps. Therefore
this identification holds with respect to a fixed basis for the 
spinor module. However, all expressions appearing in the 
Lagrangian and in the supersymmetry variations are covariant
with respect to Lorentz transformations,
since all spinor and other indices are properly contracted. Therefore
any result obtained in our distinguished coordinate system holds
in all coordinate systems.

Indices on Dirac spinors are raised and lowered using $A$ and
$A^{-1}$:
\[
\lambda^\alpha = A^{\alpha \beta} \lambda_\beta \Leftrightarrow
\lambda_\alpha = A_{\alpha \beta} \lambda^\beta \;.
\]
Spinor indices on doubled spinors $(\lambda^i_\alpha) \in \mathbb{S}
\otimes \mathbb{C}^2$ are raised and lowered using $C$ and $C^{-1}$:
\[
\lambda^{i\alpha} = C^{\alpha \beta} \lambda^i_{\beta}
\Leftrightarrow
\lambda^{i}_{\alpha} = C_{\alpha \beta} \lambda^{i \beta} \;.
\]
Internal indices $i,j=1,2$ on doubled spinors are raised and lowered
according to
\[
\lambda^i = \varepsilon^{ij} \lambda_j 
\Leftrightarrow 
\lambda_i = \lambda^j \varepsilon_{ji} \;,
\]
where $\varepsilon_{ij}=-\varepsilon_{ji}$ and
$\varepsilon^{ij}\varepsilon_{kj} = \delta^i_k= - \varepsilon^{ij}
\varepsilon_{jk}$. Note that
$(\varepsilon^{ij})=-(\varepsilon_{ij})^{-1}$, which makes raising and
lowering $SU(2)$ indices consistent with the NW-SE convention. 
In contrast, our convention for raising and lowering spinor indices 
$\alpha, \beta, \ldots$ does not comply with the NW-SE convention.

As an application we extract the structure constants of the
supersymmetry algebra from the associated vector-valued 
sesquilinear form. The natural vector-valued sesquilinear form on the spinor
module is
\[
A(\gamma^\mu \lambda, \chi) = \lambda_\alpha^* 
(\gamma^\dagger)^{\mu \alpha}_{\;\;\;\;\beta} A^{\beta \gamma}
\chi_\gamma=
\lambda^{*\alpha} \gamma^{\mu\;\;\beta}_{\;\;\alpha} A_{\beta\gamma}
\chi^\beta \;.
\]
Therefore the matrix representing $A(\gamma^\mu\cdot, \cdot)$ with
respect to the dual (`upper index')  coordinates $\lambda^\alpha, \chi^\beta$ is
\[
K[A]^\mu_{\;\;\alpha \beta} = (\gamma^\mu A^{-1})_{\alpha \beta} \;.
\]
The vector-valued real bilinear form defining the supersymmetry
algebra in terms of Dirac spinors is the real or imaginary part of
$A(\gamma^\mu\cdot, \cdot)$, depending on signature. 
Let us take the case where it is the real part, for definiteness. 
The anticommutator of a general linear combination of supercharges
is defined using the vector-valued bilinear form by
\[
\{ \lambda^\alpha Q_\alpha \;, \chi^\beta Q_\beta \}  =
A(\gamma^\mu \lambda, \chi) P_\mu = 
\lambda^\alpha \chi^\beta \mbox{Re} (\gamma^\mu A^{-1})_{\alpha \beta} P_\mu\;,
\]
from which we obtain
\[
\{ Q_\alpha \;, Q_\beta\} = \mbox{Re} (\gamma^\mu A^{-1})_{\alpha
  \beta} P_\mu \varepsilon_{ij}\;.
\]
Similarly, the bilinear form on the doubled spinor module is
\[
C(\gamma^\mu \lambda,\chi) = -\frac{1}{2} 
\lambda^i_{\alpha} (\gamma^{\mu T})^\alpha_{\;\;\beta} C^{\beta\gamma}
\chi^j_{\gamma}\varepsilon_{ji} =-\frac{1}{2} 
\lambda^{i\alpha} \chi^{j\beta} (\gamma^\mu C^{-1})_{\alpha \beta}
\varepsilon_{ij}\;.
\]
Substituting this into
\[
\{ \lambda^{i\alpha} Q_{i\alpha}\;, \chi^{j\beta} Q_{j\beta} \} = 
C(\gamma^\mu \lambda, \chi) P_\mu
\]
we obtain
\[
\{  Q_{i\alpha}\;, Q_{j\beta} \} = -\frac{1}{2} (\gamma^\mu
C^{-1})_{\alpha \beta} P_\mu \varepsilon_{ij}\;.
\]

\section{Para-quaternions \label{App-para-quaternions}}

The para-quaternions $q\in \mathbb{H}'$ are defined
by 
\[
\mathbb{H}' = \{ 
q= q_0 + q_1 e_1 + q_2 e_2 + q_3 e_{12} | q_0, q_1, q_2, q_3 \in
\mathbb{R} \} \;,
\]
where
\[
e_1^2=e_2^2=1 \;,\;\;\;e_{12}: = e_1 e_2 =- e_2 e_1 \Rightarrow
e_{12}^2=-1 \;.
\]
They form a four-dimensional real associative algebra. One defines
the conjugate
\[
q^* = q_0 - q_1 e_1 - q_2 e_2 - q_3 e_{12}
\]
and the norm
\[
N(q) = qq^* = q_0^2 - q_1^2 - q_2^2 + q_3^2 \;.
\]
A para-quaternion is invertible iff its norm is non-zero, with inverse
\[
q^{-1} = \frac{q^*}{N(q)} \;.
\]
Note that in contrast to the quaternions, the para-quaternions do not
form a skew-field. As a normed algebra, $\mathbb{H}'$ is isomorphic
to algebra $\mathbb{R}(2)$ of real $2\times 2$ matrices,  
with the norm provided by the determinant.
An isomorphism is given by
\[
1 \mapsto \left( \begin{array}{cc}
1 & 0 \\ 0 & 1 \\ \end{array} \right) \;,\;\;\;
e_1 \mapsto \left( \begin{array}{cc}
0 & 1 \\ 1 & 0 \\ \end{array} \right) \;,\;\;\;
e_2 \mapsto \left( \begin{array}{cc}
1 & 0 \\ 0 & -1 \\ \end{array} \right) \;,\;\;\;
e_{12} \mapsto \left( \begin{array}{cc}
0 & -1 \\ 1 & 0 \\ \end{array} \right) \;,
\]
so that
\[
q \mapsto M(q) = 
\left( \begin{array}{cc}
q_0 + q_2 & q_1 - q_3 \\
q_1 + q_3 & q_0 - q_2 \\
\end{array} \right) \;.
\]
Note that $\det M(q) = qq^* = N(q)$. 

Therefore the group $(\mathbb{H}')^*$ of invertible para-quaternions is
\[
(\mathbb{H}')^*:= GL(1,\mathbb{H}')  := \{ q\in \mathbb{H} | N(q) \not=0\} \simeq
GL(2,\mathbb{R}) \;,
\]
and the subgroup of unit norm para-quaternions is
\[
U(1,\mathbb{H}') := \{ q \in \mathbb{H}^* | N(q) = 1\} 
\simeq SL(2,\mathbb{R})\;.
\]
Para-quaternions can be viewed as pairs of complex numbers,
\[
q =( q_0 + e_{12} q_3) + e_1 (q_1 + e_{12} q_2) =: u + e_1 v \;,
\]
where $u,v$ are interpreted as complex numbers by identifying
$e_{12} \simeq i$.

The normed algebra $\mathbb{H}'$ can be represented by complex 
$2\times 2$ matrices:
\[
q\mapsto \tilde{M}(q)= \left( \begin{array}{cc}
u & v \\ v^* & u^* \\ \end{array} \right)  \in  \mathbb{C}(2) \;.
\]
Note that 
$N(q) = uu^* - vv^* = \det\tilde{M}$. 
The subgroup of unit norm matrices is $SU(1,1)\simeq SL(2,\mathbb{R})
\simeq U(1,\mathbb{H}')$.

Matrices of the form
$\tilde{M}(q)$ 
are invariant under the real structure on $\mathbb{C}(2)$
defined by
\[
\rho\;: \tilde{M} \mapsto \eta \tilde{M}^* \eta \;,\;\;\;
\eta = \left( \begin{array}{cc} 
0 & 1 \\ 1 & 0 \\
\end{array} \right) \;.
\]
Thus matrices of the form $\tilde{M}(q)$ form a subalgebra
isomorphic to $\mathbb{R}(2)$, as they must:
\[
\mathbb{H}' \simeq 
\{ \tilde{M} \in \mathbb{C}(2)
 | \rho(\tilde{M}) =
\tilde{M}  \} \simeq \mathbb{R}(2) \;.
\]
Further note that
$\mathbb{H}'\simeq Cl_{1,1} \simeq Cl_{0,2}$ as real associative algebras.

\section{Quaternions \label{App-quaternions}} 

For comparison let us also review standard results about the
quaternions $\mathbb{H}$. The quaternions $q\in \mathbb{H}$ are
defined by
\[
\mathbb{H} = \{ q = q_0 + q_1 i + q_2 j + q_3 k | q_0, q_1, q_2, q_3
\in \mathbb{R} \} \;,
\]
where
\[
i^2 = j^2 = -1 \;,\;\;\; k=ij = - ji \Rightarrow k^2 = -1 \;.
\]
They form a four-dimensional real associative algebra, One defines the conjugate
\[
q^* = q_0 - q_1 i - q_2 j - q_3 k
\]
and the (reduced\footnote{For the quaternion one often refers to $||q||:= N(q)^{1/2}$
  as the norm.}) norm 
\[
N(q) = qq^* = q_0^2 + q_1^2 + q_2^2 + q_3^2 \;.
\]
A quaternion is invertible iff
\[
N(q) \not= 0 \Leftrightarrow q\not= 0 \;,
\]
with inverse
\[
q^{-1} = \frac{q^*}{N(q)} \;.
\]
Since only $q=0$ does not have an inverse, the quaternions do not only
form an associative algebra, but a skew field. As a normed algebra,
$\mathbb{H}$ is isomorphic to the matrix algebra
\[
\left\{ \left( \begin{array}{cc} 
u & v \\
- v^* & u^* \\
\end{array} \right) | u,v\in \mathbb{C} \right\} \subset
\mathbb{C}(2) \;.
\]
An isomorphism is provided by
\begin{eqnarray*}
q &\mapsto& M(q) = q_0 \dblone + q_1 i\sigma^1 + q_2 i \sigma^2 + q_3 i
\sigma^3 = \left( \begin{array}{cc}
q_0 + i q_3 & q_2 + i q_1 \\
- q_2 + i q_1 & q_0 - i q_3 \\
\end{array} \right) \\
& =& \left( \begin{array}{cc}
u & v \\ - u^* & v^* \\
\end{array} \right) \;,
\end{eqnarray*}
where $\sigma^i$, $i=1,2,3$ are the Pauli matrices and where 
$u=q_0+iq_3$ and $v=q_2+iq_1$. The norm of $q$ is equal to
the determinant of the corresponding complex $2\times 2$ matrix:
\[
N(q) = a^2 + b^2 + c^2 + d^2 = uu^* + vv^* = \det M(q) \;.
\]
If we impose unit norm, then
\[
N(q) = uu^* + vv^* = 1 \;,
\]
and $M(q) \in SU(2)$. Therefore the group of unit norm quaternions is
\[
U(1,\mathbb{H}) = \{ q | N(q)=1 \} \simeq SU(2)  \;.
\]
Any matrix $M(q)$ with $N(q)\not=0$ can be written as a positive scalar
multiple of an $SU(2)$ matrix, and therefore the group of invertible
quaternions is
\[
\mathbb{H}^* = GL(1,\mathbb{H}) = \{ q \in \mathbb{H} | N(q)\not=0\} 
= \mathbb{R}^{>0} \times SU(2) \;.
\]
Further note that
$\mathbb{H} \simeq C_{2,0}$ as real associative algebras.

\end{appendix}

\providecommand{\href}[2]{#2}\begingroup\raggedright\endgroup


\end{document}